\documentclass{article}

\usepackage{PRIMEarxiv}

\usepackage[utf8]{inputenc} 
\usepackage[T1]{fontenc}    
\usepackage{hyperref}       
\usepackage{url}            
\usepackage{booktabs}       
\usepackage{amsfonts}       
\usepackage{nicefrac}       
\usepackage{microtype}      
\usepackage{lipsum}
\usepackage{fancyhdr}       
\usepackage{graphicx}       
\graphicspath{{media/}}     
\usepackage{amssymb}
\usepackage{amsmath}

\usepackage{cite}

\title{Modeling Directional Hardening and Intrinsic Size Effects Using a Dislocation Density-Based Strain Gradient Plasticity Framework
}

\author{
  Anirban Patra, Namit Pai, Parhitosh Sharma \\
  Department of Metallurgical Engineering and Materials Science \\
  Indian Institute of Technology Bombay \\
  Mumbai, India - 400076\\
  \texttt{\{Anirban Patra\}anirbanpatra@iitb.ac.in} \\
}

\begin{document}
\maketitle

\begin{abstract}
This work proposes a dislocation density-based strain gradient $J_2$ plasticity framework that models the strength contribution due to Geometrically Necessary Dislocations (GNDs) using a lower order, Taylor hardening backstress model. An anisotropy factor is introduced to phenomenologically represent the differential hardening between grains in this $J_2$ plasticity framework. An implicit numerical algorithm is implemented for the time integration of the finite deformation plasticity model. The framework is first used to predict directional hardening due to the GND-induced backstress during cyclic loading. Deformation contours are studied to understand the substructure attributes contributing to directional hardening. The framework is then used to predict the intrinsic, grain size-dependent strengthening of polygrain ensembles. Model predictions of simulations with different grain sizes are shown to agree with the Hall-Petch effect and also with Ashby's model of hardening due to GNDs in polygrain ensembles.
\end{abstract}


\section{Introduction}
\label{sec:sample1}

Plastic deformation in metallic systems is an inherently length scale-dependent phenomenon. This is generally attributed to the formation of Geometrically Necessary Dislocations (GNDs), which accommodate the strain gradients arising due to discontinuities and heterogeneities in the microstructure and the specimen during incompatible plastic deformation \cite{nye1953some, ashby1970deformation}. GNDs govern the development of long range backstress and ensuing length scale-dependent strengthening phenomena. These effects are accentuated as the size decreases, with effects on the mechanical property most evident at micron and sub-micron length scales. Physical manifestations of these length scale/size effects are evident in the Hall-Petch effect \cite{hall1951deformation, petch1953cleavage}, torsion of thin wires \cite{fleck1994strain}, bending of thin foils \cite{stolken1998microbend}, micropillar compression \cite{uchic2004sample}, nanoindentation of crystalline materials \cite{elmustafa2003nanoindentation, zhao2003material} and constrained shear of films \cite{mu2014thickness, mu2017measuring}. Further, it should be noted that the size effects on mechanical properties have generally been classified into two categories: intrinsic and extrinsic effects. While the former owe their strength to the microstructural attributes of the material, such as grain size, the latter owe their strength to the specimen size itself \cite{geers2006size, greer2011plasticity}. 

Modeling frameworks have attempted to capture these length scale-dependent effects via the development of non-local models, which account for the spatial gradient of strain/stress and their contribution to hardening during plastic deformation. In this regard, the Nye tensor was proposed as a function of the gradient of plastic strain \cite{nye1953some}. This concept has been explored and mathematically interpreted in various forms to account for the development of strain gradients, GNDs and associated hardening \cite{arsenlis1999crystallographic, gao1999mechanism, acharya2000lattice, gurtin2002gradient, acharya2006size, geers2006second, han2007finite, kysar2007high, guruprasad2008phenomenological, lele2008small, mayeur2011dislocation, dunne2012crystal, maziere2015strain, martinez2016fracture, demiral2017enhanced, gudmundson2019isotropic, kuroda2019simple, zhou2020predictive}. Recently, grain size-dependent strengthening has also predicted using strain gradient crystal plasticity models \cite{sun2019strain, haouala2020simulation, berbenni2020fast}. There has been extensive research in this field over the last three decades. Summarizing the features of these modeling studies is perhaps beyond the scope of the present work and the reader is referred to \cite{mayeur2014comparison, voyiadjis2019strain} for a comprehensive review of the research done in this field.

The focus of the present work is on modeling the development of backstress due to GNDs and their effect on the size-dependent mechanical properties. In this regard, various backstress formulations have been proposed \cite{gurtin2002gradient, evers2004non, yefimov2005size, bayley2006comparison, kuroda2006studies, forest2008some, mayeur2011dislocation} that generally account for hardening contributions due to \textit{higher order} gradient terms, based on thermodynamic considerations. More recently, Sangid and co-authors \cite{kapoor2018incorporating, bandyopadhyay2021comparative} have proposed a \textit{lower order}, GND density based backstress model for kinematic hardening in a crystal plasticity framework, such that the slip system-level backstress scales as the square root of the GND density (similar to a Taylor hardening model \cite{taylor1934mechanism}). While these recent studies \cite{kapoor2018incorporating, bandyopadhyay2021comparative} primarily focused on studying the GND and backstress evolution during cyclic loading, size effects on the mechanical properties were not studied extensively. 

In the present work, we explore this concept further and propose a $J_2$ plasticity extension of the lower order, Taylor hardening backstress model given by Kapoor et al. \cite{kapoor2018incorporating} in a dislocation density based framework. We also introduce an \textit{anisotropy factor} to phenomenologically represent the differential hardening between grains with different crystallographic orientations, in our otherwise isotropic $J_2$ plasticity model. Simulations are first performed to study the directional hardening effects due to the GND-induced backstress. Simulations are then performed to demonstrate the grain size-dependent hardening of polygrain ensembles with varying grain sizes. Based on the analysis of our model predictions, we establish correlations between the GND density, backstress evolution and grain size-dependent deformation.

\section{Model Description}
\subsection{Finite Deformation Kinematics}
\label{kinematics}
This finite deformation framework is based on the multiplicative decomposition of the deformation gradient into the elastic and plastic parts \cite{khan1995continuum}: 
\begin{equation}
\mathbf{F} = \mathbf{F}^e \cdot \mathbf{F}^p
\end{equation}
where, $\mathbf{F}^p$ relates the reference configuration to a stress-free, intermediate configuration and accounts for shear due to plastic deformation, while $\mathbf{F}^e$ relates the intermediate configuration to the current, deformed deformation and accounts for the elastic deformation.

The plastic deformation gradient is related to the velocity gradient as, $\mathbf{\dot{F}}^p = \mathbf{L}^p \cdot \mathbf{F}^p$, where the velocity gradient, $\mathbf{L}^p$, is given by \cite{weber1990finite}:
\begin{equation}
    \mathbf{L}^p = \sqrt{\frac{3}{2}}\dot{\bar{\epsilon}}^p \mathbf{N}^p
\end{equation}
Here, $\dot{\bar{\epsilon}}^p$ is the effective plastic strain rate and $\mathbf{N}^p$ is the direction of the plastic flow, given by the following relation:
\begin{equation}
   \mathbf{N}^p =  \sqrt{\frac{3}{2}}\frac{\mathbf{S} - \mathbf{\chi}}{\bar{\sigma^*}}
\end{equation}
where, $\mathbf{S}$ is the deviatoric stress tensor, $\mathbf{\chi}$ is the backstress tensor, and $\bar{\sigma^*}$ is the modified effective stress, defined as 
\begin{equation}
    \bar{\sigma^*} = \sqrt{\frac{3}{2} (\mathbf{S} - \mathbf{\chi}) : (\mathbf{S} - \mathbf{\chi})} 
\end{equation}

\subsection{Kinetics and Substructure Evolution}
\label{kinetics}
The effective plastic strain rate is modeled using a Kocks-type, thermally activated flow rule \cite{kocks1975thermodynamics}:
\begin{equation}
    \dot{\bar{\epsilon}}^p = \dot{\bar{\epsilon}}^p_0\exp\left(\frac{-\Delta F_g}{kT}\left(1 - \left(\frac{\bar{\sigma^*} - S_a}{S_t}\right)^p\right)^q\right); \bar{\sigma} > S_a
\label{eqn:flow_rule}
\end{equation}
where, $\dot{\bar{\epsilon}}^p_0$ is the reference strain rate, $\Delta F_g$ is the activation energy for dislocation glide, $S_a$ is the athermal slip resistance, $S_t$ is the thermal slip resistance, and $p$ and $q$ are parameters used to model the shape of the activation enthalpy curve. The athermal slip resistance, $S_a$, is given by:
\begin{equation}
    S_a = M(\tau_0 + k_{IH}Gb\sqrt{\rho_{SSD}})
\label{eqn:s_a}
\end{equation}
where, $\tau_0$ is threshold slip resistance, $k_{IH}$ is the Taylor hardening coefficient \cite{taylor1934mechanism} associated with isotropic hardening, $G$ is the shear modulus, $b$ is the Burgers vector magnitude, and $\rho_{SSD}$ is the Statistically Stored Dislocation (SSD) density. Here, we introduce an anisotropy factor, $M$, to phenomenologically represent the anisotropy in the yield stress of crystalline materials due to their crystallographic orientations. This term may be considered to be representative of the Taylor factor, or the inverse of the Schmid factor, in this macroplasticity framework and is randomly assigned for each grain. As a first order approximation, $M$ is assumed to be constant and its evolution with plastic deformation is neglected. We also note that the anisotropy factor may be expected to evolve with deformation (cf. \cite{mishra2019new}).

The evolution of SSDs is modeled using a Kocks-Mecking-Estrin type formulation \cite{estrin1996dislocation, kocks2003physics}, \textit{albeit} with the additional consideration for GNDs in the generation of SSDs \cite{evers2004non}:
\begin{equation}
\dot{\rho}_{SSD} = \frac{k_{mul}}{b}\sqrt{\rho_{SSD} + \rho_{GND}}\dot{\bar{\epsilon}}^p - k_{rec}\rho_{SSD}\dot{\bar{\epsilon}}^p
\label{eqn:rho_ssd}
\end{equation}
where, $k_{mul}$ is the dislocation multiplication rate constant, and $k_{rec}$ is the recovery constant. The first term in Equation (\ref{eqn:rho_ssd}) represents the multiplication of SSDs at existing dislocations, while the second term represents the rate of annihilation of SSDs due to recovery processes.

\subsection{Geometrically Necessary Dislocation Density and Backstress Tensor}
In this finite deformation framework, we adopt the definition of the Nye tensor proposed by Dai \cite{dai1997geometrically} as
\begin{equation}
    \mathbf{\Lambda} = - (\mathbf{\nabla} \times \mathbf{F}^{pT})^T
\label{eqn:nye}
\end{equation}
Further, the Geometrically Necessary Dislocation (GND) density is defined as
\begin{equation}
    \rho_{GND} = \frac{1}{b} || \mathbf{\Lambda} ||
\end{equation}
where, $|| ||$ denotes the L2 norm of the respective tensor quantity.

In the present work, we implement Equation (\ref{eqn:nye}) in its rate form, such that
\begin{equation}
    \mathbf{\dot \Lambda} = - (\mathbf{\nabla} \times \mathbf{\dot F}^{pT})^T; \dot \rho_{GND} = \frac{1}{b} || \mathbf{\dot \Lambda} ||
\label{eqn:nye_dot}    
\end{equation}
Further, we propose the backstress tensor and its rate form as
\begin{equation}
    \mathbf{\chi} = k_{KH}Gb\sqrt{\rho_{GND}} \mathbf{N}^p \Leftrightarrow \mathbf{\dot \chi} = k_{KH}Gb \frac{\dot \rho_{GND}}{2 \sqrt{\rho_{GND}}} \mathbf{N}^p  
\end{equation}
where, $k_{KH}$ is the Taylor hardening coefficient associated with kinetic hardening due to GNDs. 

This form of the backstress tensor is a $J_2$ plasticity extension of the Taylor hardening contribution of the GND density to the slip system-level directional hardening proposed for a crystal plasticity framework \cite{kapoor2018incorporating}. Physically, this model represents the development of backstress due to GNDs along the direction of plastic deformation, $\mathbf{N}^p$. The rate (or the incremental) form of this backstress model ensures that the backstress contribution due to the pile up of GND density along a certain loading direction is remnant in the history-dependent backstress tensor, even after the direction of loading is changed, for example, during cyclic loading. In the context of prior macroplasticity frameworks, our formulation is a deviation from the mechanism-based strain gradient plasticity models that proposed the hardening due to the total density of SSDs and GNDs (cf. \cite{nix1998indentation, gao1999mechanism} and its variants). Essentially, our model delineates the contribution of the GNDs towards the development of backstress in the direction of plastic deformation, while the SSDs are assumed to contribute primarily to isotropic hardening. Note, however, that higher order strain gradient frameworks do separate the strengthening contribution of SSDs and GNDs into isotropic hardening and kinematic hardening, respectively \cite{evers2004non, evers2004scale, bayley2006comparison}, although their physical implications are different. 

\subsection{Implicit Time Integration and Numerical Implementation}
We propose an implicit algorithm for the time integration of the constitutive model. This algorithm is inspired by the implicit slip rate integration algorithms \cite{cuitino1993computational, mcginty2001multiscale, mcginty2006semi, ling2005numerical}, which were given for crystal plasticity formulations. Here we adapt the same for a $J_2$ plasticity formulation. 

Given the deformation gradient, $\mathbf{F}$, at any time step, this algorithm numerically estimates the effective plastic strain rate, $\dot{\bar{\epsilon}}^p$, via a Newton-Raphson algorithm. At any given time step, a function may be formulated such that the function, $f(\dot{\bar{\epsilon}}^p)$, given by:
\begin{equation}
    f(\dot{\bar{\epsilon}}^p)_{i+1} = f(\dot{\bar{\epsilon}}^p)_{i} + \frac{\partial{f(\dot{\bar{\epsilon}}^p)}}{\partial{\dot{\bar{\epsilon}}^p}}\Delta\dot{\bar{\epsilon}}^p 
\end{equation}
has to be minimized iteratively, i.e, $f \rightarrow 0$. Here, the subscript, $i$, denotes the number of iterations at any given time step. In order to satisfy this equality,
\begin{equation}
    f(\dot{\bar{\epsilon}}^p)_{i} = -\frac{\partial{f(\dot{\bar{\epsilon}}^p)}}{\partial{\dot{\bar{\epsilon}}^p}}\Delta\dot{\bar{\epsilon}}^p \Leftrightarrow \Delta\dot{\bar{\epsilon}}^p = - \frac{f(\dot{\bar{\epsilon}}^p)_{i}}{\frac{\partial{f(\dot{\bar{\epsilon}}^p)}}{\partial{\dot{\bar{\epsilon}}^p}}}
\label{eqn:NReqn}
\end{equation}
Thus, $\dot{\bar{\epsilon}}^p_{i+1} = \dot{\bar{\epsilon}}^p_{i} + \Delta\dot{\bar{\epsilon}}^p$ provides the converged value of the effective plastic strain rate at any given time step.

This method relies on accurate calculation of the partial derivatives associated with the $\frac{\partial{f(\dot{\bar{\epsilon}}^p)}}{\partial{\dot{\bar{\epsilon}}^p}}$ term. Taking derivatives and expanding the individual terms gives an expression of the form:
\begin{equation}
\begin{split}
    \frac{\partial{f(\dot{\bar{\epsilon}}^p)}}{\partial{\dot{\bar{\epsilon}}^p}} = 1 - \dot{\bar{\epsilon}}^p\left(q\frac{\Delta{F_g}}{kT}\left(1-\left(\frac{\bar{\sigma^*} - S_a}{S_t}\right)^p\right)^{q-1}\right)\left(p\left(\frac{\bar{\sigma^*} - S_a}{S_t}\right)^{p-1}\right) \\
    \left(\frac{1}{S_t}\left(\frac{\partial{\bar{\sigma^*}}}{\partial{\dot{\bar{\epsilon}}^p}} - \frac{\partial{S_a}}{\partial{\dot{\bar{\epsilon}}^p}}\right) - \left(\frac{\bar{\sigma^*} - S_a}{S_t^2}\right)\frac{\partial{S_t}}{\partial{\dot{\bar{\epsilon}}^p}}\right)
\end{split}
\end{equation}
The above expression contains derivatives of various terms with respect to $\dot{\bar{\epsilon}}^p$, which are described in the following. We begin with $\frac{\partial{\bar{\sigma^*}}}{\partial{\dot{\bar{\epsilon}}^p}}$, which may be written as \cite{mcginty2001multiscale}
\begin{equation}
    \frac{\partial{\bar{\sigma^*}}}{\partial{\dot{\bar{\epsilon}}^p}} = \frac{\partial{\bar{\sigma^*}}}{\partial{\mathbf{\sigma}}} : \frac{\partial{\mathbf{\sigma}}}{\partial{\mathbf{\sigma}^{pk2}}} : \frac{\partial{\mathbf{\sigma}^{pk2}}}{\partial{\mathbf{E}^e}} : \frac{\partial{\mathbf{E}^e}}{\partial{\mathbf{F}^p}} : \frac{\partial{\mathbf{F}^p}}{\partial{\mathbf{L}^p}} : \frac{\partial{\mathbf{L}^p}}{\partial{\dot{\bar{\epsilon}}^p}}
\end{equation}
Here, $\mathbf{\sigma}$ is the Cauchy stress, $\mathbf{\sigma}^{pk2}$ is the second Piola-Kirchoff stress, and $\mathbf{E}^e$ is the elastic Green strain. Following McGinty \cite{mcginty2001multiscale}, this expression may be simplified as
\begin{equation}
    \frac{\partial{\bar{\sigma^*}}}{\partial{\dot{\bar{\epsilon}}^p}} \approx - \sqrt{\frac{3}{2}} \mathbf{N}^p : \mathbf{C} : \sqrt{\frac{3}{2}} \mathbf{N}^p
\end{equation}
where, $\mathbf{C}$ is the fourth rank elasticity tensor. Further, the derivative of the athermal stress with respect to the effective plastic strain rate is given as 
\begin{equation}
    \frac{\partial{S_a}}{\partial{\dot{\bar{\epsilon}}^p}} = \frac{k_{IH}Gb}{2\sqrt{\rho_{SSD}}}\frac{\partial{\rho_{SSD}}}{\partial{\dot{\bar{\epsilon}}^p}}
\end{equation}
where,
\begin{equation}
   \frac{\partial{\rho_{SSD}}}{\partial{\dot{\bar{\epsilon}}^p}} = \frac{\left(\frac{k_{mul}}{b}\sqrt{\rho_{SSD} + \rho_{GND}} - k_{dyn}\rho_{SSD} + \frac{k_{mul}}{2b\sqrt{\rho_{SSD} + \rho_{GND}}}\frac{\partial\rho_{GND}}{\partial{\dot{\bar{\epsilon}}^p}}\right)\Delta t}{1 - \frac{k_{mul}}{2b\sqrt{\rho_{SSD} + \rho_{GND}}}\dot{\bar{\epsilon}}^p\Delta t + k_{dyn}\dot{\bar{\epsilon}}^p\Delta t} 
\end{equation}
and,
\begin{equation}
   \frac{\partial{\rho_{GND}}}{\partial{\dot{\bar{\epsilon}}^p}} = \frac{1}{b} ||\frac{\partial\mathbf{\dot \Lambda}}{\partial{\dot{\bar{\epsilon}}^p}}|| {\Delta t}    
\end{equation}
Here,
\begin{equation}
    \frac{\partial\mathbf{\dot \Lambda}}{\partial{\dot{\bar{\epsilon}}^p}} = - \left( \mathbf{\nabla} \times \frac{\partial \mathbf{\dot F}^p}{\partial{\dot{\bar{\epsilon}}^p}} ^T \right)^T
\end{equation}
Calculating the derivative of $\mathbf{\dot F^p}$ with respect to $\dot{\bar{\epsilon}}^p$ is not straightforward. Following McGinty \cite{mcginty2001multiscale}, we write $\mathbf{F^p}$ at time step $t + \Delta t$ as:
\begin{equation}
    \mathbf{F}^p_{t+\Delta t} = exp(\mathbf{L}^p_0 \Delta t) \cdot \mathbf{F}^p_t
\label{eqn:Fp_t}    
\end{equation}
where, $\mathbf{L^p_0}$ is the plastic velocity gradient in the intermediate configuration. This expression may be expanded using the Cayley Hamilton theorem \cite{curtis1990algebraic, mcginty2001multiscale}, i.e,
\begin{equation}
    exp(\mathbf{L}^p_0 \Delta t) = \mathbf{I} + \frac{sin \phi}{\phi} \mathbf{L}^p_0 \Delta t + \frac{1-cos \phi}{\phi^2}(\mathbf{L}^p_0 \cdot \mathbf{L}^p_0) {\Delta t}^2
\end{equation}
where, 
\begin{equation}
    \phi = \left(\sqrt{\frac{1}{2} \mathbf{L}^p_0 : \mathbf{L}^p_0} \right) \Delta t
\end{equation}
In order to calculate $\mathbf{\dot F}^p$, we take the finite difference of Equation (\ref{eqn:Fp_t}) with respect to $\Delta t$ such that
\begin{equation}
    \mathbf{\dot F}^p = \frac{\mathbf{F}^p_{t+\Delta t} - \mathbf{F}^p_{t}}{\Delta t} = \frac{1}{\Delta t} \left(exp(\mathbf{L}^p_0 \Delta t) - \mathbf{I} \right) \cdot \mathbf{F}^p_t
\end{equation}
Using the expression for $exp (\mathbf{L}^p_0 \Delta t)$ given above and after some mathematical manipulation, we arrive at 
\begin{equation}
\begin{split}
    \frac{\partial \mathbf{\dot F}^p}{\partial{\dot{\bar{\epsilon}}^p}} = \frac{1}{\Delta t} \left(\frac{\partial (exp (\mathbf{L}^p_0 \Delta t))}{\partial{\dot{\bar{\epsilon}}^p}} \cdot \mathbf{L}^{p-1}_0 \cdot \mathbf{\dot F}^p + (exp (\mathbf{L}^p_0 \Delta t) - \mathbf{I}) \cdot \frac{\partial \mathbf{L}^{p-1}_0}{\partial{\dot{\bar{\epsilon}}^p}} \cdot \mathbf{\dot F}^p \right) \cdot \\ 
    \left(\mathbf{I} - \frac{1}{\Delta t} (exp (\mathbf{L}^p_0 \Delta t) - \mathbf{I}) \cdot \mathbf{L}^{p-1}_0 \right) ^{-1}
\end{split}    
\end{equation}
Further, the individual derivatives in the above expression are given as:
\begin{equation}
\begin{split}
    \frac{\partial (exp (\mathbf{L}^p_0 \Delta t))}{\partial{\dot{\bar{\epsilon}}^p}} = \frac{sin \phi}{\phi} \frac{\partial \mathbf{L}^p_0}{\partial{\dot{\bar{\epsilon}}^p}} \Delta t + \frac{1 - cos \phi}{\phi^2} \left(\frac{\partial \mathbf{L}^p_0}{\partial{\dot{\bar{\epsilon}}^p}} \cdot \mathbf{L}^p_0 + \mathbf{L}^p_0 \cdot \frac{\partial \mathbf{L}^p_0}{\partial{\dot{\bar{\epsilon}}^p}} \right) \Delta t^2 \\ 
    + \left(\frac{cos \phi}{\phi} - \frac{sin \phi}{\phi ^2} \right) \frac{\partial \phi}{\partial{\dot{\bar{\epsilon}}^p}} (\mathbf{L}^p_0 \Delta t) + \left(\frac{\sin \phi}{\phi^2} - \frac{1 - cos \phi}{2 \phi^3} \right) \frac{\partial \phi}{\partial{\dot{\bar{\epsilon}}^p}} \mathbf{L}^p_0 \cdot \mathbf{L}^p_0 \Delta t^2
\end{split}    
\end{equation}
\begin{equation}
    \frac{\partial \phi}{\partial{\dot{\bar{\epsilon}}^p}} = \frac{\Delta t}{2 \sqrt{\frac{1}{2} \mathbf{L}^p_0 : \mathbf{L}^p_0}} \frac{\partial \mathbf{L}^p_0}{\partial{\dot{\bar{\epsilon}}^p}} : \mathbf{L}^p_0
\end{equation}
\begin{equation}
    \frac{\partial \mathbf{L}^p_0}{\partial{\dot{\bar{\epsilon}}^p}} = \sqrt{\frac{3}{2}} \mathbf{N}^p
\end{equation}
\begin{equation}
    \frac{\partial \mathbf{L}^{p-1}_0}{\partial{\dot{\bar{\epsilon}}^p}} = - \mathbf{L}^{p-1}_0 \cdot \frac{\partial \mathbf{L}^p_0}{\partial{\dot{\bar{\epsilon}}^p}} \cdot \mathbf{L}^{p-1}_0
  \label{eqn:dLP_inv}    
\end{equation}
Here, Equation (\ref{eqn:dLP_inv}) is derived using linear algebra concepts \cite{petersen2008matrix}. The above set of equations provide all the terms required for the implicit Newton-Raphson algorithm. We solve Equation (\ref{eqn:NReqn}) iteratively until the change in the effective plastic strain rate, $\Delta\dot{\bar{\epsilon}}^p$, is below a prescribed tolerance.

The constitutive model has been implemented as a user-defined material model and interfaced with the open source finite element (FE) code, Multiphysics Object-Oriented Simulation Environment (MOOSE) \cite{permann2020moose}. MOOSE provides the material model with an increment of the deformation gradient at an integration point, based on the applied global boundary conditions. The material model performs an implicit update of the stress, internal state variables and the consistent tangent stiffness tensor at the Gauss point, which is then passed back to MOOSE to check for global convergence. While the FE framework considers the displacement variables as degrees of freedom, the spatial gradient terms are implemented using the \textit{auxiliary variable} interface in MOOSE. Note that these \textit{auxiliary variables}, computed at the Gauss points, are not used in the computation of the global Jacobian, hence reducing computation costs significantly. In our implementation, $\mathbf{\dot F}^{p}$ is stored as an \textit{auxiliary variable} and its spatial gradient calculated using standard finite element shape functions of desired order for determining the rate of Nye tensor (cf. Equation (\ref{eqn:nye_dot})).

\section{Model Parameters and Simulation Details}
\label{sec:parameters}
The model parameters chosen in this work are given in Table \ref{tab:params}. In this example application, we have not calibrated our model to the mechanical response of any specific material and have used parameters that provide a \textit{representative} stress-strain response and flow strength expected of metallic systems.

The elastic constants and Burgers vector magnitude are chosen to be representative of a fcc metal. The Kocks-type thermally activated flow rule has been extensively used in crystal plasticity frameworks to model the temperature- and strain rate-dependent response of various metals and alloys \cite{kothari1998elasto, ranjan2021crystal, sedighiani2021determination} and our choice of flow rule parameters is inspired from these earlier studies. The Taylor hardening coefficient associated with isotropic hardening due to SSDs is chosen to be $0.2$ \cite{taylor1934mechanism}. Since the focus of this work is to highlight the contribution of GNDs to the strengthening, we have intentionally chosen a low value of the initial SSD density as $1.0 \times 10^5 mm^{-2}$. Further, the dislocation multiplication constant for SSDs is also chosen to have a low value of $0.01$, such that the SSD density does not increase significantly during deformation. 

As a first order approximation, the initial GND density has been assumed to be zero. This may considered as representative of an annealed material, which has no residual deformation. In the present work, $k_{KH}$ is the most important parameter governing the development of backstress and its contribution to size-dependent strengthening. In the following sections, we have demonstrated model predictions with two different values for this parameter: $1.2$ and $0.8$. In Section \ref{sec:scaling_relations}, we have also provided a rationale for choosing appropriate values for this parameter.

\begin{table}[!htbp]
  \begin{center}
    \caption{Constitutive model parameters.}
    \label{tab:params}
    \begin{tabular}{c c c}
     \hline
       Parameter &  Value & Meaning \\ \hline
       $E$ & 128 GPa & Young's modulus \\
       $\nu$ & $0.34$ & Poisson's ratio \\
       $G$ & 47.76 GPa & Shear modulus \\
       $b$ & $2.56 \times 10^{-10} m$ & Burgers vector magnitude \\
       $\dot{\bar{\epsilon}}^p_0$ & $1.0 \times 10^{-2} s^{-1}$ & Reference strain rate \\
       $\Delta F_g$ & $4G{b}^3$ & Activation energy for dislocation glide \\
       $S_t$ & 400 MPa & Thermal slip resistance \\
       $\tau_0$ & 50 MPa & Threshold slip resistance \\
       $k_{IH}$ & 0.2 & Isotropic hardening coefficient due to SSDs \\
       $k_{KH}$ & 1.2 and 0.8 & Kinematic hardening coefficient due to GNDs \\
       $k_{mul}$ & 0.01 & Dislocation multiplication rate constant \\
       $k_{rec}$ & 500 & Dislocation recovery constant \\
       $\rho_{SSD}^0$ & $1.0 \times 10^5 mm^{-2}$ & Initial SSD density \\
       $\rho_{GND}^0$ & $0$ & Initial GND density \\
       \hline
    \end{tabular}
  \end{center}
\end{table}

We have performed simulations of polygrain ensembles with varying grain sizes. The initial microstructures were instantiated using the open source Voronoi tessellation software, Neper \cite{quey2011large}, for a 2D simulation domain of $100 \times 100 \mu m$. The microstructures were meshed using 2D six node, triangular finite elements with quadratic interpolation, and a mesh size of $\approx 1.5 \mu m$. There were approximately $10,000-20,000$ finite elements in each of these microstructures. The microstructures generated by Neper were converted to the exodus file format (required for MOOSE) using the commercial meshing software, Trelis \cite{trelisadvanced}. Further, the anisotropy factor, $M$, was randomly assigned for each grain, such that it ranges from $1 \slash 0.27$ to $1 \slash 0.49$. Here, $0.27$ and $0.49$ are representative of the lower and upper bounds of the Schmid factor generally observed in cubic crystals deforming by octahedral slip. As mentioned earlier, $M$ is not allowed to evolve during the simulation. The primary purpose of this parameter is to introduce intergranular heterogeneity in the $J_2$ plasticity simulations.

Axi-symmetric boundary conditions were used such that the left edge of the simulation domain was constrained in the x-direction, while the bottom edge was constrained in the y-direction. The corner node common to both these edges was constrained in all degrees of freedom to prevent rigid body motion. Displacement-controlled uniaxial loading was applied on the top edge at a nominal strain rate of $1.0 \times 10^{-4} s^{-1}$.

\begin{figure}[!htbp]
	\centering
	\includegraphics[scale=0.4]{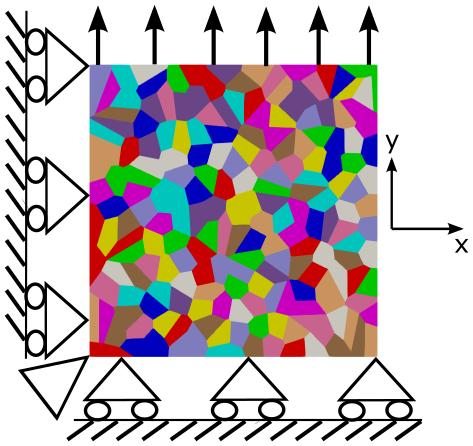}
	\caption{Representative microstructure with an average grain size of $8 \mu m$ in the simulation domain of $100 \times 100 \mu m^2$, along with a schematic of the loading and boundary conditions.} 
	\label{fig:bcs}
\end{figure}

A schematic of the loading and boundary conditions along with a representative microstructure is shown in Figure \ref{fig:bcs}.

\section{Results and Discussion}

\subsection{Directional Hardening Effects}
We first present model predictions of directional hardening. We have used the microstructure with $8 \mu m$ average grain size for these simulations. The microstructure was cyclically loaded in tension and compression up to $0.01$ strain amplitude for $500 s$ at a nominal strain rate of $1.0 \times 10^{-4} s^{-1}$. In order to demonstrate the effect of $k_{KH}$ on directional hardening, we performed three simulations for the same initial microstructure. While the first two simulations were performed with $k_{KH} = 1.2$ and $k_{KH} = 0.8$, the third simulation was performed with $k_{KH} = 0$ and using a local version of the $J_2$ plasticity model. Essentially, all the strain gradient-dependent terms, $\rho_{GND}$ and $\mathbf{\chi}$ were assumed to be zero in this third simulation.

\subsubsection{Aggregate Properties}
\begin{figure}[!htbp]
	\centering
	\includegraphics[scale=0.25]{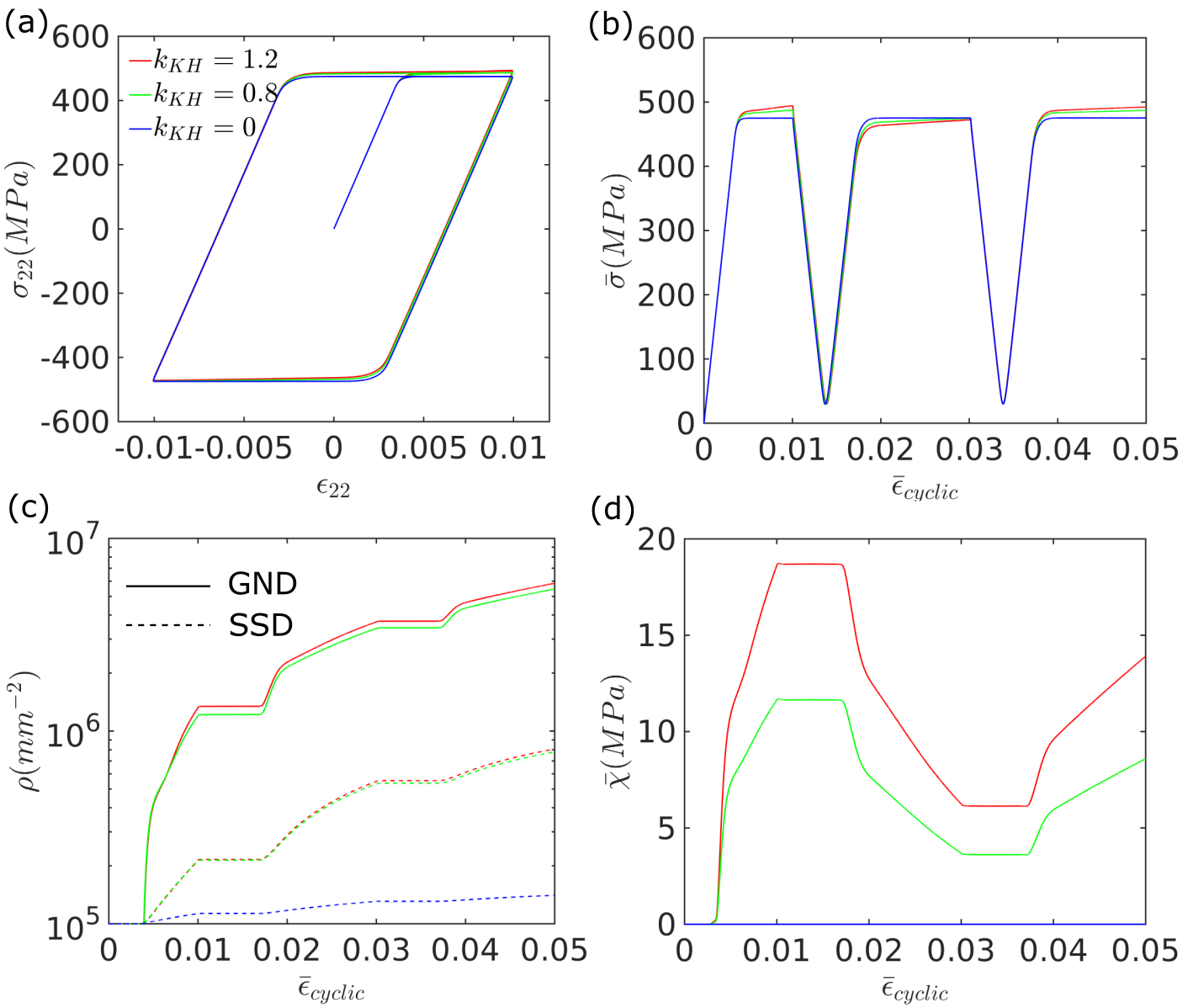}
	\caption{Plot of (a) $\sigma_{22}$ versus $\epsilon_{22}$, (b) $\bar \sigma$ versus $\bar \epsilon_{cyclic}$, (c) average $\rho_{GND}$ and $\rho_{SSD}$ versus $\bar \epsilon_{cyclic}$, and (d) $\bar \chi$ versus $\bar \epsilon_{cyclic}$ for simulations with different values of $k_{KH}$. In (c), solid lines are used to denote $\rho_{GND}$, while dotted lines are used to denote $\rho_{SSD}$. Note that the same color scheme has been used to represent the different values of $k_{KH}$ in all the plots.} 
	\label{fig:cyclic_aggregate}
\end{figure}

Figure \ref{fig:cyclic_aggregate} (a) and (b) show the aggregate response, plotted in terms of $\sigma_{22}$ versus $\epsilon_{22}$, and the effective stress, $\bar \sigma$, versus effective cyclic strain, $\bar \epsilon_{cyclic}$, for simulations with different values of $k_{KH}$. It can be seen that the flow stress increases with increasing $k_{KH}$ in the 1st cycle of loading. For example, $\bar \sigma$ is $494 MPa$ for $k_{KH} = 1.2$, $487 MPa$ for $k_{KH} = 0.8$, and $475 MPa$ for $k_{KH} = 0$ after $\bar \epsilon_{cyclic} = 0.01$. Subsequent to tension, when the microstructures were loaded in compression, the flow stress for the simulation with $k_{KH} = 0$ is almost the same as that in tension, indicating only isotropic hardening. For $k_{KH} = 0.8$ and $1.2$, the flow stresses in compression are initially lower, followed by subsequent hardening. As discussed earlier, model parameters related to isotropic hardening due to SSDs were intentionally kept low in our simulations. Hence, the hardening observed in our simulations is primarily due to the GND-dominated kinematic hardening. The net hardening due to GND density may appear to be relatively small. For example, the hardening after $\bar \epsilon_{cyclic} = 0.01$ is $\approx 19 MPa$ using $k_{KH} = 1.2$ and $\approx 12 MPa$ using $k_{KH} = 0.8$, as compared to the simulations with $k_{KH} = 0$. However, it will be shown in the later sections that this size-dependent strength contribution is consistent with the values expected of fcc metals. Further, the value of the parameter, $k_{KH}$, may simply be increased, if a higher directional hardening is desired.

Figure \ref{fig:cyclic_aggregate} (c) shows the evolution of the average $\rho_{GND}$ and $\rho_{SSD}$ as a function of $\bar \epsilon_{cyclic}$ for the three cases shown in Figure \ref{fig:cyclic_aggregate} (a). The GND density is shown using solid lines, while the SSD density is shown using dotted lines. It can be seen that the GND density is slightly higher for $k_{KH} = 1.2$, as compared to that for $k_{KH} = 0.8$. While the strain gradient itself is expected to be similar in both cases, some additional GNDs may have developed to accommodate the stress concentrations due to higher backstress. This will be evident in the deformation contours discussed in the next section. Further, the SSD density is similar for both cases, while it is 5-7 times higher than the case with $k_{KH} = 0$ (no strain gradient). Since the model parameters for SSD evolution have been intentionally kept low, we do not see significant evolution in the SSD density for the case with no strain gradient. For $k_{KH} = 1.2$ and $0.8$, there is some evolution of SSDs, which may be primarily attributed to the multiplication of SSDs at GND segments, modeled using the 1st term in Equation (\ref{eqn:rho_ssd}). 

Before moving forward, we would like to point out that the sign of the GND density is not considered in our $J_2$ plasticity framework. For example, GNDs with opposite sign may be expected to develop during reverse loading and hence contribute to a reduction in the net GND density. However, the GND density is computed from the \textit{norm} of the Nye tensor in our model (cf. Equation (\ref{eqn:nye_dot})) and hence there is no way to account for the sign of the GNDs in our framework. Nonetheless, the backstress tensor accounts for the directionality of the hardening caused due the GND density. This is evident from the evolution of the effective backstress, $\bar \chi$, with $\bar \epsilon_{cyclic}$, in Figure \ref{fig:cyclic_aggregate}(d), which shows that the effective backstress decreases once plastic deformation commences during compression loading, and then increases again during tension loading. This is physically representative of the hardening contribution due to backstress changing signs, once the loading direction is reversed. It should also be noted that the backstress scales almost linearly with the parameter $k_{KH}$. For example, the average value of $\bar \chi$ is $18.68 MPa$ after $\bar \epsilon_{cyclic} = 0.01$ using $k_{KH} = 1.2$, while it is $11.67 MPa$ using $k_{KH} = 0.8$, which gives a scaling of $\approx 2 \slash 3$. The GND density and backstress are of course zero using $k_{KH} = 0$.

In summary, the results presented in this section show the capability of the model to predict directional hardening and backstress evolution under cyclic loading conditions. While the GND density is not signed in this $J_2$ plasticity framework, the backstress tensor appropriately accounts for the directional hardening effects due to the GND density.

\subsubsection{Deformation and Substructure Contours}
\begin{figure}[!htbp]
	\centering
	\includegraphics[scale=0.22]{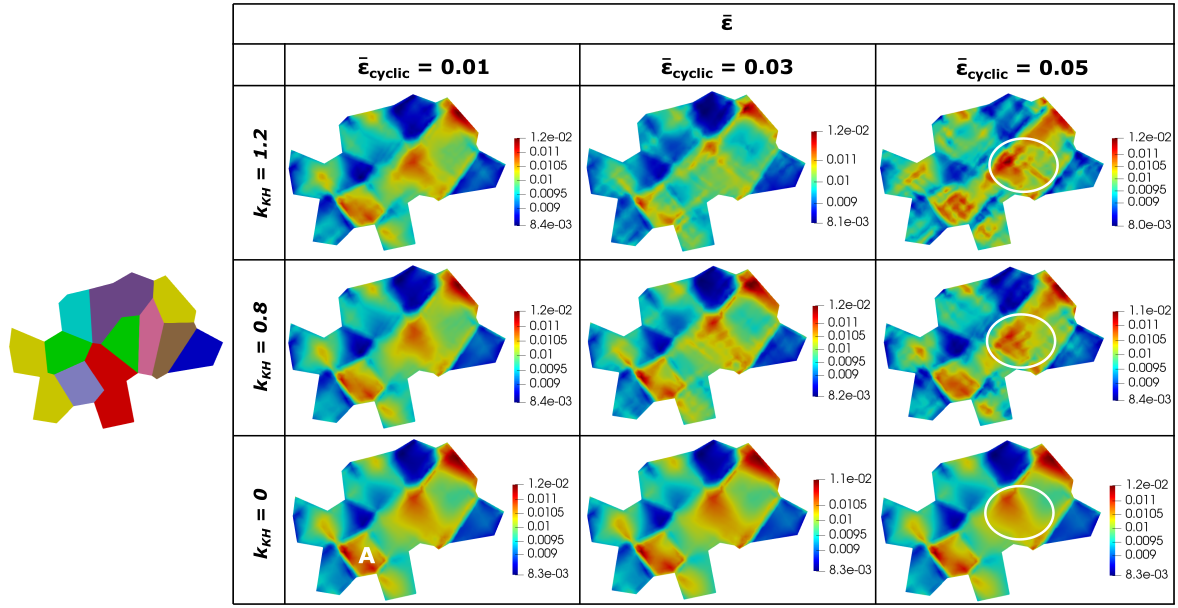}
	\caption{Contours of effective strain, $\bar \epsilon$, at the end of tension loading during the 1st cycle ($\bar \epsilon_{cyclic} = 0.01$), the end of compression loading during the 1st cycle ($\bar \epsilon_{cyclic} = 0.03$), and at the end of tension loading during the 2nd cycle ($\bar \epsilon_{cyclic} = 0.05$) for simulations with different values of $k_{KH}$. The set of grains chosen for these contours is shown on the left side.} 
	\label{fig:cyclic_strain_contours}
\end{figure}

\begin{figure}[!htbp]
	\centering
	\includegraphics[scale=0.25]{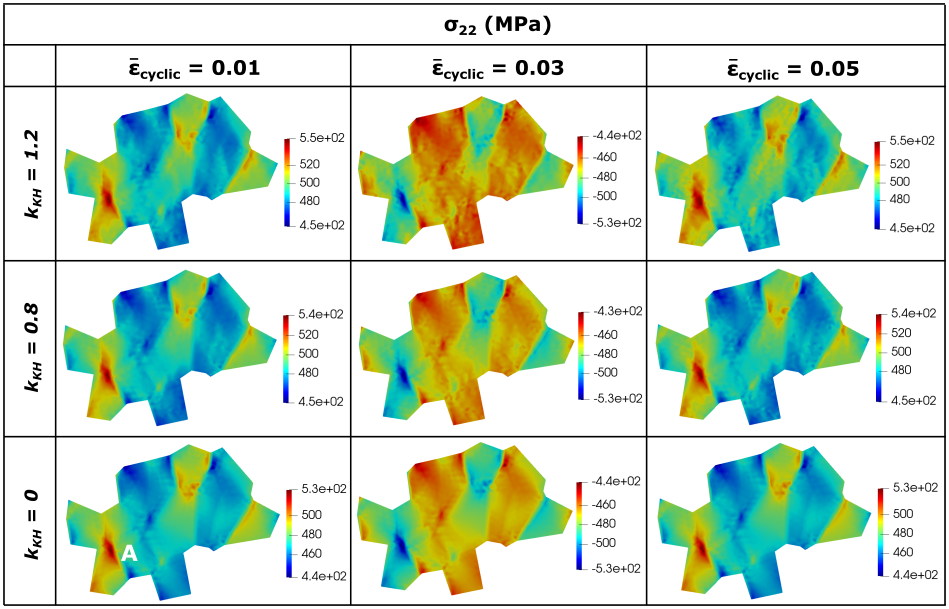}
	\caption{Contours of $\sigma_{22}$ at the end of tension loading during the 1st cycle ($\bar \epsilon_{cyclic} = 0.01$), the end of compression loading during the 1st cycle ($\bar \epsilon_{cyclic} = 0.03$), and at the end of tension loading during the 2nd cycle ($\bar \epsilon_{cyclic} = 0.05$) for simulations with different values of $k_{KH}$. The set of grains chosen for these contours is shown on the left side of Figure \ref{fig:cyclic_strain_contours}.} 
	\label{fig:cyclic_stress_contours}
\end{figure}

In order to highlight the ability of the model to predict heterogeneous deformation, we present deformation contours from the cyclic loading simulations in this section. For this, we choose a small set of grains from the initial microstructure shown in Figure \ref{fig:bcs}. These grains are highlighted on the left side of Figure \ref{fig:cyclic_strain_contours}. The contours of effective strain, $\bar \epsilon$, at different stages of cyclic loading are shown for the three cases in Figure \ref{fig:cyclic_strain_contours}. Essentially, we have plotted the strain contours at the end of tension loading during the 1st cycle ($\bar \epsilon_{cyclic} = 0.01$), the end of compression loading during the 1st cycle ($\bar \epsilon_{cyclic} = 0.03$), and at the end of tension loading during the 2nd cycle ($\bar \epsilon_{cyclic} = 0.05$). The corresponding contours of $\sigma_{22}$ are shown in Figure \ref{fig:cyclic_stress_contours}. The effective strain is expected to be similar at these three stages of deformation.

Firstly, it is interesting to note that the model predicts heterogeneous deformation in the different grains using this $J_2$ plasticity framework. This can be clearly seen from the $\bar \epsilon$ and $\sigma_{22}$ contours, where certain grains have higher strains, while other grains have higher stresses. Generally, such effects can be captured using crystal plasticity models. In our model, this is attributed to the introduction of the anisotropy factor, $M$ (cf. Equation \ref{eqn:s_a}), which is used to phenomenologically represent the differential hardening between grains. This also promotes strain and stress concentrations at the interfaces between grains, for example in the region indicated by A in Figures \ref{fig:cyclic_strain_contours} and \ref{fig:cyclic_stress_contours}. It should also be noted that such effects are observed in all cases, with and without strain gradients.

The effect of strain gradients on directional hardening can be clearly observed in the strain contours for the different cases at $\bar \epsilon_{cyclic} = 0.05$. This is highlighted with circular markers in Figure \ref{fig:cyclic_strain_contours}. In the absence of strain gradients and directional hardening ($k_{KH} = 0$), the only heterogeneity in deformation may arise from the differential hardening between grains, due to the anisotropy factor, $M$. However, in the presence of strain gradients and with increasing values of $k_{KH}$, the strain heterogeneities increase in the same regions. These strain heterogeneities also increase with applied deformation, with early precursors of strain localization being observed at $\bar \epsilon_{cyclic} = 0.05$. 

The $\sigma_{22}$ contours in Figure \ref{fig:cyclic_stress_contours} show that with increasing $k_{KH}$, the stresses are not fully reversed when the loading is reversed from tension to compression and vice-versa. For example, the maximum and minimum values of $\sigma_{22}$ are $\approx 550 MPa$ and $\approx 450 MPa$ at the end of the tension loading for $k_{KH} = 1.2$, while the same at the end of the compression cycle are $\approx -440 MPa$ and $\approx -530 MPa$, respectively. For $k_{KH} = 0$ and no strain gradients, the corresponding absolute values are identical between tension and compression. Interestingly, the grains with higher relative stress in tension have lower relative stress in compression, and vice-versa.

\begin{figure}[!htbp]
	\centering
	\includegraphics[scale=0.25]{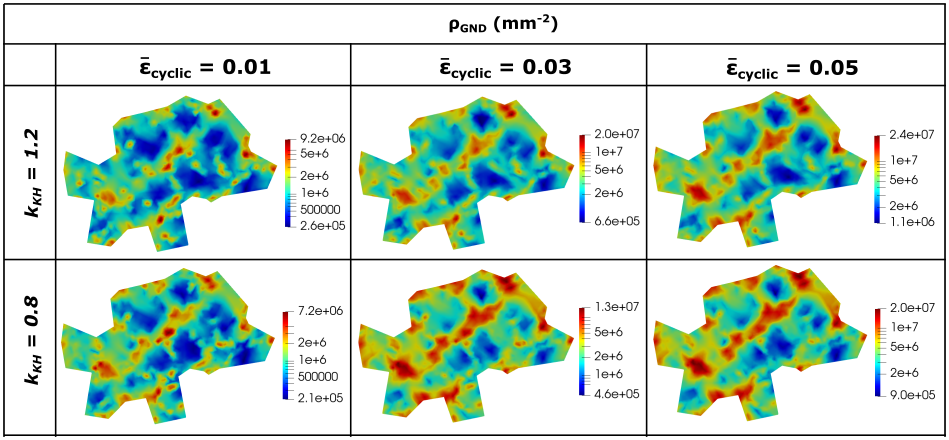}
	\caption{Contours of $\rho_{GND}$ at the end of tension loading during the 1st cycle ($\bar \epsilon_{cyclic} = 0.01$), the end of compression loading during the 1st cycle ($\bar \epsilon_{cyclic} = 0.03$), and at the end of tension loading during the 2nd cycle ($\bar \epsilon_{cyclic} = 0.05$) for simulations with different values of $k_{KH}$. The set of grains chosen for these contours is shown on the left side of Figure \ref{fig:cyclic_strain_contours}.} 
	\label{fig:cyclic_gnd_contours}
\end{figure}

\begin{figure}[!htbp]
	\centering
	\includegraphics[scale=0.25]{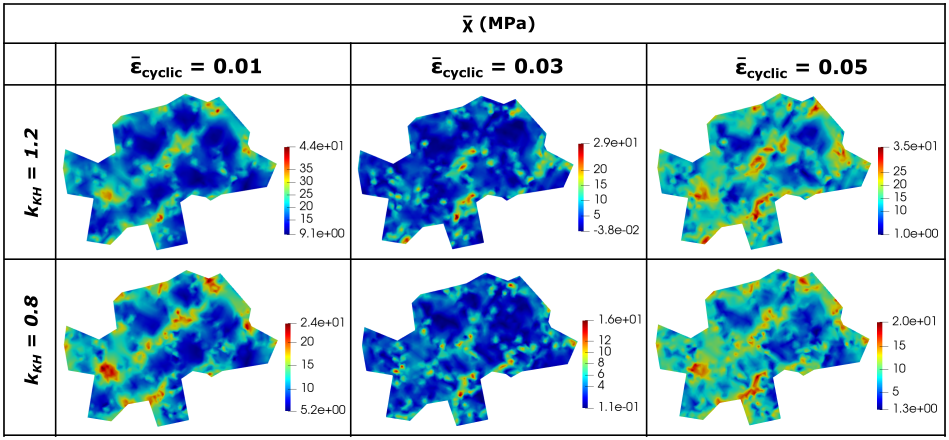}
	\caption{Contours of $\bar \chi$ at the end of tension loading during the 1st cycle ($\bar \epsilon_{cyclic} = 0.01$), the end of compression loading during the 1st cycle ($\bar \epsilon_{cyclic} = 0.03$), and at the end of tension loading during the 2nd cycle ($\bar \epsilon_{cyclic} = 0.05$) for simulations with different values of $k_{KH}$. The set of grains chosen for these contours is shown on the left side of Figure \ref{fig:cyclic_strain_contours}.} 
	\label{fig:cyclic_chi_contours}
\end{figure}

Figure \ref{fig:cyclic_gnd_contours} shows contours of the GND density, $\rho_{GND}$, at different stages of applied deformation for $k_{KH} = 1.2$ and $k_{KH} = 0.8$, while Figure \ref{fig:cyclic_chi_contours} shows contours of the effective backstress, $\bar \chi$, for the same. As can be seen, the GND density mostly develops at the grain interfaces and triple junctions, which are regions of stress concentration and incompatible deformation between grains. Further, the GND density intensifies at these regions with increasing applied deformation. As mentioned earlier, our $J_2$ model does not account for the sign of the GND density and although the GND density may increase with applied deformation, their net effect may be to reduce the backstress. The same can be seen from the effective backstress contours in Figure \ref{fig:cyclic_chi_contours}, where the effective backstress generally reduces during the compression cycle in all the grains. In fact, the maximum and minimum values of $\bar \chi$ at the end of tension loading during the 2nd cycle ($\bar \epsilon_{cyclic} = 0.05$) are lower than the corresponding values after the end of tension loading during the 1st cycle ($\bar \epsilon_{cyclic} = 0.01$). This is primarily due to the fact that the overall backstress hardening during compression loading negates the same developed during tension loading in the 1st cycle.

The results presented in this section demonstrate the ability of the framework to simulate directional hardening due to GNDs. In the following section, we focus on establishing size effects using our model.

\subsection{Grain Size Effects}
In order to establish the grain size-dependent strengthening relations, we have performed simulations with five different mean grain sizes: $4 \mu m$, $8 \mu m$, $20 \mu m$, $30 \mu m$, and $40 \mu m$. Essentially, we span across an order of magnitude in terms of the mean grain size. Note that while this represents the mean grain size provided to Neper \cite{quey2011large}, there is some variation in the actual grain sizes obtained during the tessellations. Further, the domain size was kept constant at $100 \times 100 \mu m$ for all simulations. All simulations were performed till $0.04$ applied strain for the material loaded in uniaxial tension. We first present the aggregate properties, followed by the deformation contours and finally estimate the grain size-dependent scaling relations.

\subsubsection{Aggregate Properties}
\begin{figure}[!htbp]
	\centering
	\includegraphics[scale=0.25]{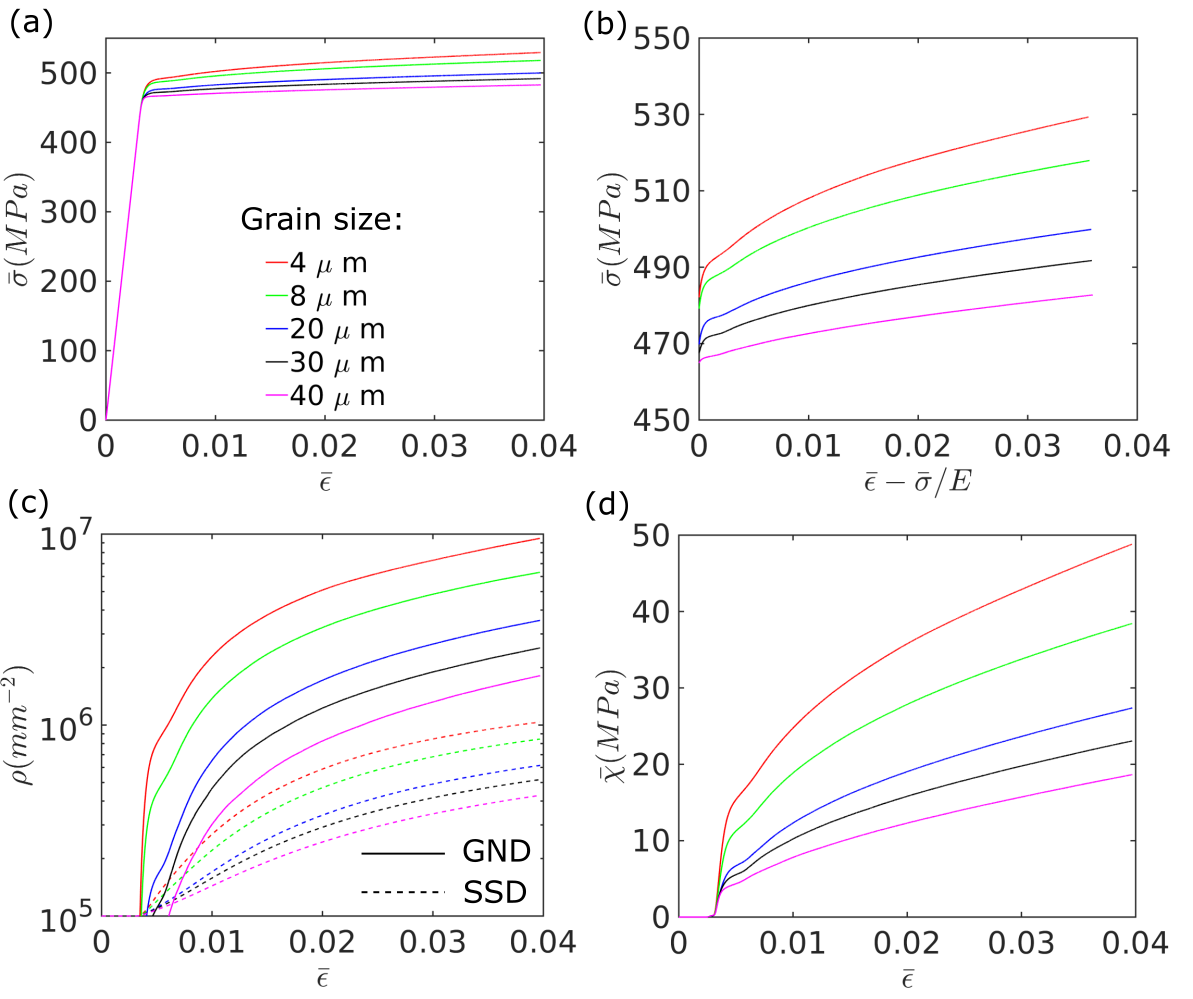}
	\caption{Plot of (a) $\bar \sigma$ versus $\bar \epsilon$, (b) $\bar \sigma$ versus $\bar \epsilon^p \approx \bar \epsilon - \bar \sigma /E$, (c) average $\rho_{GND}$ and $\rho_{SSD}$ versus $\bar \epsilon$, and (d) $\bar \chi$ versus $\bar \epsilon$ for simulations with different grain sizes. In (c), solid lines are used to denote $\rho_{GND}$, while dotted lines are used to denote $\rho_{SSD}$. Note that the same color scheme has been used to represent the different grain sizes in all the plots.} 
	\label{fig:aggregate_properties}
\end{figure}

Figure \ref{fig:aggregate_properties}(a) shows the aggregate response, plotted in terms of the effective stress, $\bar \sigma$, versus effective strain, $\bar \epsilon$, for simulations with different average grain sizes. In order to highlight the grain size-dependent response, the plot of $\bar \sigma$ versus $\bar \epsilon^p$ given by $\approx \bar \epsilon - \bar \sigma /E$ is shown in Figure \ref{fig:aggregate_properties}(b). The corresponding average GND and SSD densities are plotted as a function of $\bar \epsilon$ in Figure \ref{fig:aggregate_properties}(c), and the average effective backstress, $\bar \chi$, is plotted as a function of $\bar \epsilon$ in Figure \ref{fig:aggregate_properties}(d).

It can be seen from Figure \ref{fig:aggregate_properties}(a) and (b) that the flow stress of the material increases as the grain size decreases. These results indicate that the model is able to qualitatively capture the grain size-dependent strengthening \cite{hall1951deformation, cordero2016six}. For example, the flow stress at the beginning of plastic deformation is $\approx 480 MPa$ for grain size $4 \mu m$, while it is $\approx 465 MPa$ for grain size $40 \mu m$. After 0.04 applied strain, the flow stress is $\approx 530 MPa$ for grain size of $4 \mu m$, while it is $\approx 482 MPa$ for grain size of $40 \mu m$. Further, Figure \ref{fig:aggregate_properties}(c) and (d) show that the GND density and backstress initially increase at a high rate at the beginning of plastic deformation and these values tend to saturate during the later stages of deformation. An effective backstress of $\approx$ 50 MPa developed after 0.04 applied strain for grain size $4 \mu m$, while the same for grain size $40 \mu m$ is less than 20 MPa. Also note that the rate of increase of GND density and backstress is higher for the lower grain sizes and vice-versa. As mentioned in Section \ref{sec:parameters}, a low value was used for the dislocation multiplication constant, $k_{mul}$, in our simulations, which resulted in a relatively lower average SSD density as compared to the GND density. Increasing the value of this parameter may contribute to increase in SSD density and increased strain hardening during plastic deformation. These ensemble average quantities are used to derive the size-dependent scaling relations in Section \ref{sec:scaling_relations}.

\subsubsection{Deformation and Substructure Contours}
\begin{figure}[!htbp]
	\centering
	\includegraphics[scale=0.22]{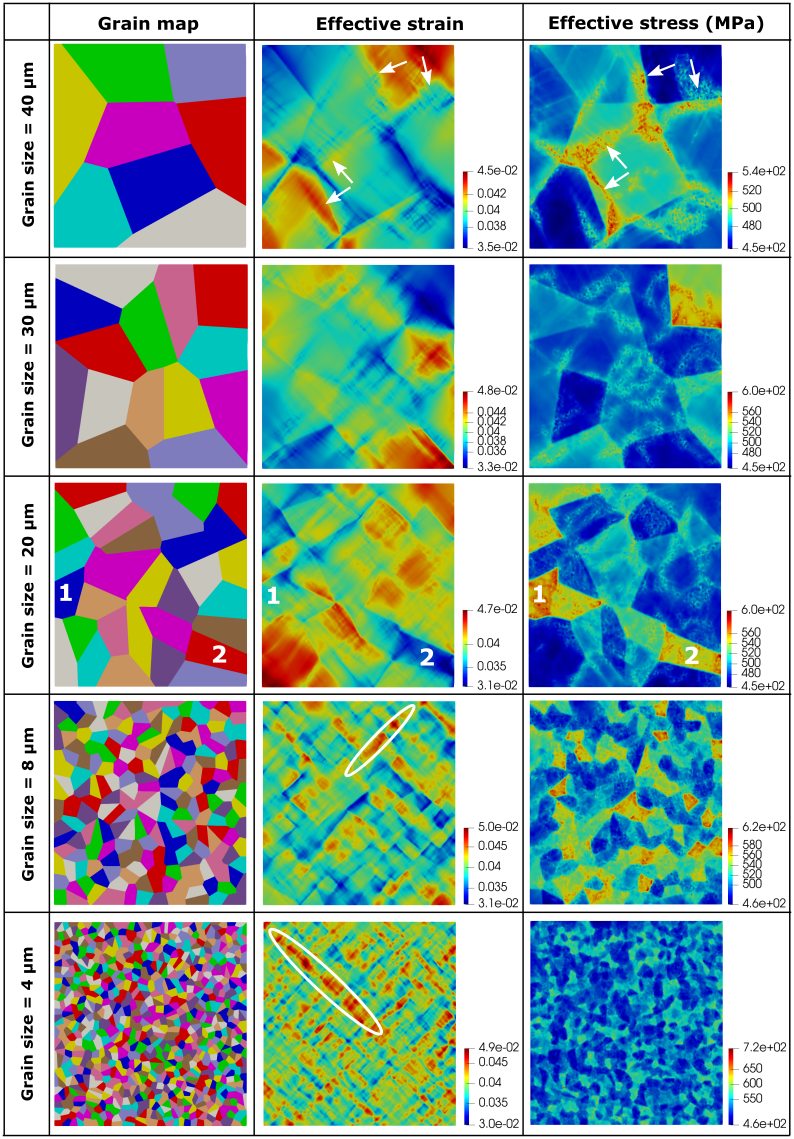}
	\caption{Grain maps and contours of effective strain, $\bar \epsilon$, and effective stress, $\bar \sigma$, after 0.04 applied strain for simulations with different grain sizes. Note that the scales are different for the variables for each grain size.} 
	\label{fig:strain_contours}
\end{figure}

\begin{figure}[!htbp]
	\centering
	\includegraphics[scale=0.22]{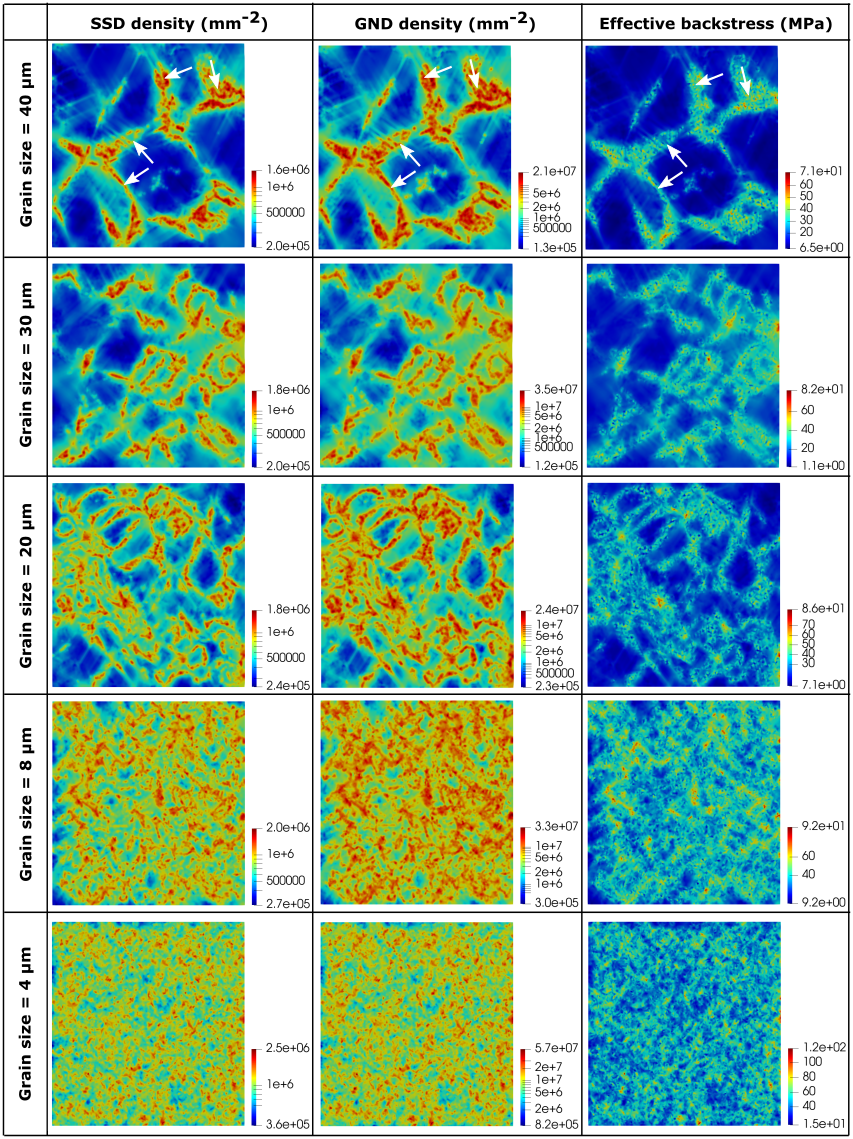}
	\caption{Contours of SSD density, $\rho_{SSD}$, GND density, $\rho_{GND}$ and effective backstress, $\bar \chi$, after 0.04 applied strain for simulations with different grain sizes. Note that the scales are different for the different variables for each grain size.}
	\label{fig:isv_contours}
\end{figure}

Figure \ref{fig:strain_contours} shows the grain maps and the corresponding contours of $\bar \epsilon$ and $\bar \sigma$ after 0.04 applied strain for the different grain sizes. The corresponding contours of $\rho_{SSD}$, $\rho_{GND}$ and $\bar \chi$ are shown in Figure \ref{fig:isv_contours}. The material was strained along the vertical direction in all cases. All deformation contours shown in this section are from simulations using $k_{KH} = 1.2$.

The effective strain contours show clear development of strain gradients at the interfaces between grains. For example, see regions marked by the white arrows in the strain and stress contour for grain size $40 \mu m$ in Figure \ref{fig:strain_contours}. The effect of strain gradients can be observed in the development of GNDs and backstress in the same regions highlighted in Figure \ref{fig:isv_contours}. Due to the increase in GND density and backstress in the vicinity of grain interfaces, higher plastic deformation occurs and the SSD density increases in these regions as well. Generally speaking, the effective strain is higher in grains with a lower effective stress and vice-versa. For example, see grains marked 1 and 2 in Figure \ref{fig:strain_contours}. As mentioned earlier, our model does not consider crystallographic orientation-dependent deformation and hardening. Rather, the anisotropy factor, $M$, is used to introduce intergranular heterogeneity in our simulations. While a crystal plasticity model may be more adept at capturing these near boundary gradient zones \cite{evers2004scale, mishra2009widths}, our $J_2$ plasticity model is still able to qualitatively simulate heterogeneous deformation near these grain interfaces that may give rise to grain size-dependent hardening.

It can be clearly seen from Figures \ref{fig:strain_contours} and \ref{fig:isv_contours} that the density of strain localizations, SSD density, GND density and backstress increase with decrease in grain size. For example, the strain contours in Figure \ref{fig:strain_contours} show precursors of deformation bands for grain sizes of $8 \mu m$ and $4 \mu m$ (see regions highlighted with ellipses). At higher applied strain, these regions may show significant strain localization. The GND density and backstress increase with decreasing grain size, as the relative fraction of grain interfaces to grain interiors increases. For example, it can be seen from Figure \ref{fig:isv_contours} that $\bar \chi$ is 5-10 times higher near the grain interfaces as compared to the grain interior for the simulation with grain size $40 \mu m$. Further, it can also be observed that the density of GNDs is higher than that of the SSDs by up to an order of magnitude near the grain interfaces, even for the simulations with high grain sizes.

Overall, it can be concluded from the contours presented in this section that our model is able to qualitatively capture deformation traits due to the intergranular heterogeneity via the introduction of the anisotropy factor, $M$, in our $J_2$ plasticity simulations. Further, the observed trends in the development of $\bar \epsilon$, $\bar \sigma$, $\rho_{SSD}$, $\rho_{GND}$ and $\bar \chi$ for simulations with different grain sizes show clear traits of grain size-dependent deformation.

\subsubsection{Grain Size-Dependent Scaling Relations}
\label{sec:scaling_relations}
The Hall-Petch equation generally used to represent \textit{intrinsic} grain size-dependent deformation is of the form \cite{hall1951deformation}:
\begin{equation}
    \sigma = \sigma_0 + \sigma_{size-dependent} = \sigma_0 + \frac{k_{HP}}{\sqrt{D}}
\label{eqn:hall_petch}    
\end{equation}
where, $\sigma$ is the flow stress, $\sigma_0$ is the strength contribution due to grain size-independent mechanisms, $\sigma_{size-dependent}$ is the grain size-dependent strength contribution, $k_{HP}$ is the Hall-Petch coefficient and $D$ is the mean grain size. $k_{HP}$ is a constant that varies for metals with different crystal structures \cite{cordero2016six}. Based on analysis of relevant experimental data, Cordero et al. \cite{cordero2016six} have also shown that $k_{HP}$ is a function of the applied strain, shear modulus and Burgers vector magnitude, i.e.,
\begin{equation}
    k_{HP} \propto G \sqrt{b \epsilon}
\end{equation}
where, $\epsilon$ is the macroscopic strain. This is based on Asbhy's model for hardening due to GNDs in polygrain ensembles \cite{ashby1970deformation}, where it was proposed that $\rho_{GND} \propto \epsilon /bD$. Note that variants of this scaling relation has also been proposed in \cite{el2015unravelling, sun2019strain, haouala2020simulation}. We try to establish these scaling relations from our model predictions here.

In our model, the strength contribution due to GNDs may be attributed solely to the backstress term. By inverting the flow rule given in Equation (\ref{eqn:flow_rule}), we arrive at the following equation for the flow stress:
\begin{equation}
    \bar \sigma^{*} = M(\tau_0 + k_{IH}Gb\sqrt{\rho_{SSD}}) + S_t \left(1 - \left(\frac{kT}{\Delta F_g} log\left(\frac{\dot{\bar{\epsilon}}^p_0}{\dot{\bar{\epsilon}}^p} \right) \right)^{1/q} \right)^{1/p}
\label{eqn:flow_stress}
\end{equation}
For the small strains and uniaxial loading conditions considered in our simulations, we further approximate Equation (\ref{eqn:flow_stress}) to a 1D form as
\begin{equation}
    \bar \sigma \approx \bar \sigma^* + \bar \chi \approx M(\tau_0 + k_{IH}Gb\sqrt{\rho_{SSD}}) + S_t \left(1 - \left(\frac{kT}{\Delta F_g} log\left(\frac{\dot{\bar{\epsilon}}^p_0}{\dot{\bar{\epsilon}}^p} \right) \right)^{1/q} \right)^{1/p} + k_{KH}Gb\sqrt{\rho_{GND}}
\label{eqn:1d_size}
\end{equation}
Here, the first two terms on the RHS represent the size-independent strength contributions and are similar to the $\sigma_0$ term in Equation (\ref{eqn:hall_petch}), while the third term on the RHS is representative of the size-dependent strength contribution due to GNDs, i.e.,
\begin{equation}
    \sigma_{size-dependent} = k_{KH}Gb\sqrt{\rho_{GND}}
\label{eqn:sig_size}
\end{equation}

We first verify that the average backstress, $\bar \chi$, obtained from our simulations is comparable to the value of $\sigma_{size-dependent}$ obtained using the average GND density (Equation (\ref{eqn:sig_size})). Further, we present results using two different values of $k_{KH}$, i.e., 1.2 and 0.8. The same initial microstructures, \textit{albeit} with random instantiations of $M$, were used. These values after 0.04 applied strain are presented in Table \ref{tab:gnd_backstress}. While there are some small differences between $\sigma_{size-dependent}$ and $\bar \chi$, these values are comparable for all cases. $\bar \chi$ is higher than $\sigma_{size-dependent}$ by $\approx 6-8 \%$ for the lower grain sizes, where stress concentrations may have developed due to the high values of backstress near the triple junctions or grain interfaces (cf. Figure \ref{fig:isv_contours}). Under such multi-axial stresses, the 1D approximation used in Equation (\ref{eqn:1d_size}) may no longer be appropriate. Nonetheless, this analysis provides verification that the strength contribution due to the backstress term indeed scales as the square root of the GND density in our model.

\begin{table}[!htbp]
  \begin{center}
    \caption{Calculation of $\sigma_{size-dependent}$ using $\rho_{GND}$ and comparison with simulated values of $\bar \chi$ after 0.04 applied strain.}
    \label{tab:gnd_backstress}
    \begin{tabular}{c c c c c c}
     \hline
       $k_{KH}$ & Grain size ($\mu m$) & $\rho_{GND}$ ($mm^{-2}$) & $\sigma_{size-dependent}$ (MPa) & $\bar{\chi}$ (MPa) \\ \hline
       1.2 & 3.93 & $9.50 \times 10^6$ & 45.03 & 48.81 \\
           & 7.93 & $6.31 \times 10^6$ & 36.69 & 38.42 \\
           & 20.02 & $3.53 \times 10^6$ & 27.46 & 27.37 \\
           & 30.21 & $2.53 \times 10^6$ & 23.26 & 23.05 \\
           & 39.94 & $1.81 \times 10^6$ & 19.67 & 18.65 \\
       0.8 & 3.93 & $8.68 \times 10^6$ & 28.70 & 30.00 \\
           & 7.93 & $5.33 \times 10^6$ & 22.50 & 22.84 \\
           & 20.02 & $2.82 \times 10^6$ & 16.37 & 15.32 \\
           & 30.21 & $1.92 \times 10^6$ & 13.50 & 12.85 \\
           & 39.94 & $1.38 \times 10^6$ & 11.43 & 10.17 \\
       \hline
    \end{tabular}
  \end{center}
\end{table}

\begin{figure}[!htbp]
	\centering
	\includegraphics[scale=0.27]{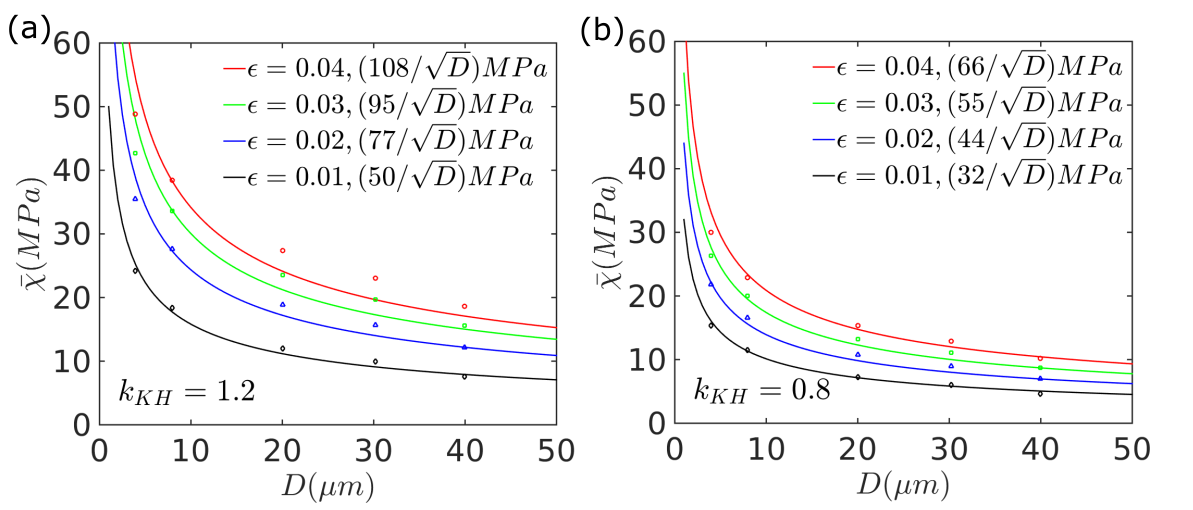}
	\caption{Plot of $\bar \chi$ versus $D$, along with the fit to the Hall-Petch equation at different applied strains for (a) $k_{KH} = 1.2$, and (b) $k_{KH} = 0.8$. The legend indicates the applied strain as well as the best fit Hall-Petch equation for each set of data.} 
	\label{fig:scaling_plots}
\end{figure}

Finally, we estimate the grain size-dependent scaling relations obtained from our simulations. For this analysis, we have fit the effective backstress, $\bar \chi$, to the mean grain size, $D$, using the Hall-Petch equation (Equation (\ref{eqn:hall_petch})) at different values of applied strain. The fit to the simulated $\bar \chi$ versus $D$ data and the corresponding Hall-Petch coefficients at different applied strains are shown in Figure \ref{fig:scaling_plots}(a) for $k_{KH} = 1.2$ and in Figure \ref{fig:scaling_plots}(b) for $k_{KH} = 0.8$. Based on the obtained fit, $k_{HP}$ ranges from $50 MPa \sqrt{\mu m}$ to $108 MPa \sqrt{\mu m}$ for the different applied strains using $k_{KH} = 1.2$. Detailed analysis of experimental data for various fcc metals \cite{cordero2016six} has shown that the Hall-Petch coefficient ranges between $70-110 MPa \sqrt{\mu m}$, except for Ni, which has $k_{HP} \approx 230 MPa \sqrt{\mu m}$. We have chosen $k_{KH}$ in our model such that the obtained Hall-Petch coefficient is in the range of experimentally observed values. Further, it should be noted that this model parameter provides almost a linear scaling of the strength contribution, manifested in terms of the backstress. For example, by reducing $k_{KH}$ from 1.2 to 0.8, the Hall-Petch coefficients at the corresponding strains scale by a factor of $\approx 2 \slash 3$. This can be seen from the range of Hall-Petch coefficients that lie between $32-66 MPa \sqrt{\mu m}$, obtained using $k_{KH} = 0.8$ in Figure \ref{fig:scaling_plots}(b). 

It is worth noting that the value of the $k_{KH}$ chosen to represent hardening due to GNDs is much higher than the corresponding value of $k_{IH}$ used to represent hardening due to SSDs. Note that there is an additional anisotropy factor, $M$, present in the hardening contribution due to SSDs. Based on the random instantiations, $M$ was generally found to be $\approx 2.5-2.6$ in our simulations. On dividing $k_{KH}$ by $M$, we get $k_{KH} \slash M \approx 0.45-0.48$. This lies at the upper bound of strength coefficient values generally used to represent Taylor hardening due to dislocations \cite{argon2008strengthening}. It should also be noted that this parameter is expected to be mesh-dependent and a smaller value of $k_{KH}$ may be needed to obtain the same strength contribution using a finer mesh, where a higher GND density may be expected \cite{bandyopadhyay2021comparative}.

\begin{figure}[!htbp]
	\centering
	\includegraphics[scale=0.3]{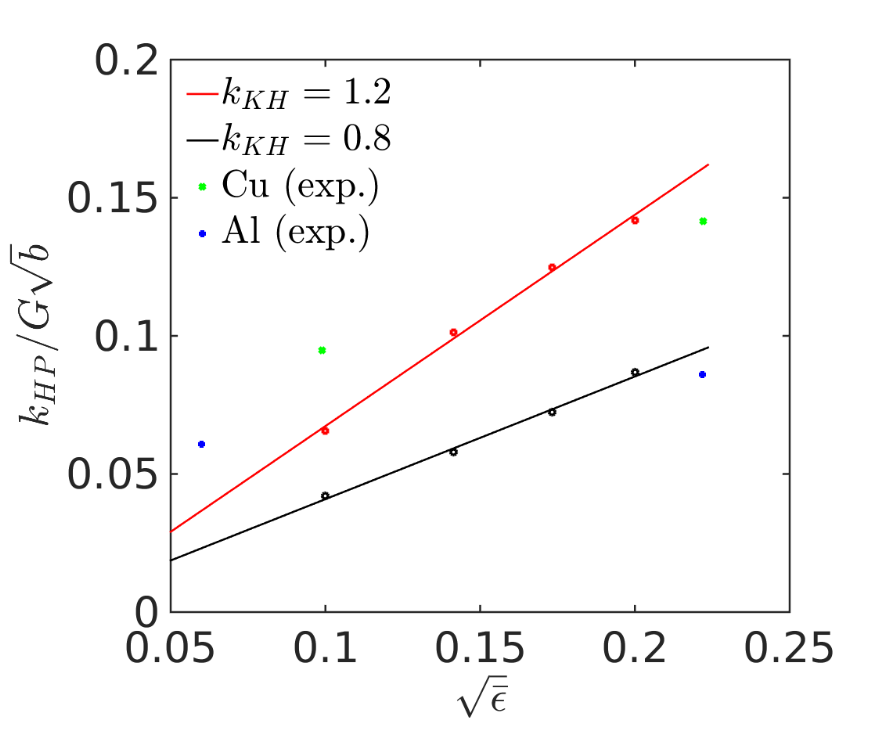}
	\caption{Plot of $k_{HP} \slash G \sqrt{b}$ versus $\sqrt{\bar \epsilon}$ as obtained from Figure \ref{fig:scaling_plots}(a) and (b), along with the best fit straight line passing through these points and comparison with experimental data for Cu and Al. The experimental data points for Cu are taken from \cite{ono1982grain, cordero2016six} and the experimental data points for Al are taken from \cite{hansen1977effect, cordero2016six}.} 
	\label{fig:ashby_plot}
\end{figure}

As mentioned earlier, Cordero et al. \cite{cordero2016six} have shown that $k_{HP} \propto G \sqrt{b \epsilon}$, by an extension of Ashby's model for GNDs in polygrain ensembles \cite{ashby1970deformation}. We have plotted the fitted values of $k_{HP} \slash G \sqrt{b}$ obtained from our simulations as a function of $\sqrt{\bar \epsilon}$ in Figure \ref{fig:ashby_plot} to verify this. The best fit straight line going through these points has also been plotted for $k_{KH} = 1.2$ and $k_{KH} = 0.8$. We have also plotted the available experimental data for Cu \cite{ono1982grain, cordero2016six} using green symbols and that for Al \cite{hansen1977effect, cordero2016six} using blue symbols in the same figure. It can be seen that the data points follow a linear trend for a given value of $k_{KH}$. Thus, our model predictions are shown to agree with the Ashby model of strengthening due to GNDs \cite{ashby1970deformation, cordero2016six}. Further, while our model parameters were not calibrated to any particular material, the values of $k_{HP} \slash G \sqrt{b}$, as a function of $\sqrt{\bar \epsilon}$, predicted by our model lie in the same range as the experimental data for Cu and Al. 

In summary, the following grain size-dependent scaling relation proposed in the literature has also been established from our model predictions:
\begin{equation}
    \sigma_{size-dependent} \propto G \sqrt{\frac{b \bar \epsilon}{D}}
\end{equation}

\subsection{Discussion}
Having presented the model and its predictions, we emphasize on the key differences between our modeling framework and a \textit{conventional} $J_2$ plasticity model here. Our framework accounts for both isotropic and kinematic hardening, which may contribute to the expansion and shifting of the yield surface, respectively. However, the isotropic hardening itself varies from grain-to-grain and is a function of the anisotropy factor, $M$, thus leading to individual and distinct yield in each of the constituent grains. Moreover, we have introduced dislocation density-based constitutive models for modeling the evolution of substructure during plastic deformation. Kinematic hardening, which contributes to the shifting of the yield surface, is assumed to be dominated by the strengthening due to GNDs. As has been shown in our results, this is able to predict both directional hardening and also grain size-dependent strengthening. While the former can also be predicted by a \textit{conventional} $J_2$ plasticity framework, the latter cannot. Moreover, the deformation-induced microstructure evolution predicted by our model is generally expected from crystal plasticity simulations. Further, it should be noted that \textit{intrinsic} grain size-dependent strengthening has generally been simulated using crystal plasticity frameworks in the literature \cite{sun2019strain, berbenni2020fast, haouala2020simulation}. Our $J_2$ plasticity simulations are expected to be significantly cheaper, in terms of computational costs, as compared to crystal plasticity simulations. 

Another discussion point worth mentioning here is that full field Crystal Plasticity Finite Element (CPFE) models are generally more computationally expensive than Fast Fourier Transform (FFT)-based spectral crystal plasticity models \cite{eisenlohr2013spectral}. This is due to the large number of degrees of freedom involved and may limit the consideration of large domains for prediction of grain size effects in crystal plasticity finite element simulations. In fact, some of the above mentioned crystal plasticity studies \cite{haouala2020simulation, berbenni2020fast} are indeed FFT-based. Our model may provide an alternative for prediction of grain size strengthening using finite element simulations, which perhaps offer more convenience for consideration of realistic geometries and imposing various realistic boundary conditions (as compared to FFT simulations).

Prediction of anisotropic hardening and grain size-dependent strengthening using this computationally efficient framework (as compared to a crystal plasticity framework) is one of the main contributions of our work. In future work, the anisotropy factor, $M$, may be allowed to evolve with deformation  \cite{mishra2019new} in order to phenomenologically represent grain rotations during polycrystalline deformation more accurately. Orientation-dependent elasticity could also be introduced to further represent anisotropic deformation in this framework.

\section{Conclusions}
We have proposed a dislocation density-based strain gradient $J_2$ plasticity model for simulating the grain size-dependent hardening response of metallic systems. In this framework, a lower order, Taylor hardening backstress model is used to represent the hardening due to GNDs. Further, an anisotropy factor, $M$, is proposed to phenomenologically represent the differential hardening between grains. An implicit numerical algorithm has been implemented for the time integration of the finite deformation plasticity model.

\begin{itemize}
    \item Our model predicted increased directional hardening during cyclic loading due to the GND-induced backstress, with higher values of the kinematic hardening parameter. The contribution of the backstress to strain heterogeneities also increased with higher values of this parameter. 
    \item Deformation contours also showed the development of strain gradients, GND density and backstress in the vicinity of grain interfaces. These effects intensified as the grain size was decreased in our simulations and are due to the introduction of the anisotropy factor, $M$, in our model.
    \item Aggregate properties predicted by our simulations were used to determine the strength scaling relations. A Hall-Petch type, grain size-dependent strength contribution \cite{hall1951deformation} was predicted by our model. 
    \item Our model predictions also agree with Ashby's model of strengthening due to GNDs in polygrain ensembles \cite{ashby1970deformation, cordero2016six}. Model predictions of $k_{HP} \slash G \sqrt{b}$, as a function of $\sqrt{\bar \epsilon}$, were found to lie in the same range as that for fcc metals from available experimental data in the literature. 
    \item Based on the analysis of grain size-dependent strengthening contributions, a rationale is also provided for choosing appropriate values of the kinematic hardening parameter to capture the Hall-Petch effect.
\end{itemize}

\section*{Acknowledgments}
The authors gratefully acknowledge funding received from the Department of Science and Technology (DST) - Science and Engineering Research Board (SERB) for this research under grant number: CRG/2020/000593.

\bibliographystyle{unsrt}  
\bibliography{SGP}

\begin{thebibliography}{10}

\bibitem{nye1953some}
John~F Nye.
\newblock Some geometrical relations in dislocated crystals.
\newblock {\em Acta metallurgica}, 1(2):153--162, 1953.

\bibitem{ashby1970deformation}
MF~Ashby.
\newblock The deformation of plastically non-homogeneous materials.
\newblock {\em The Philosophical Magazine: A Journal of Theoretical
  Experimental and Applied Physics}, 21(170):399--424, 1970.

\bibitem{hall1951deformation}
EO~Hall.
\newblock The deformation and ageing of mild steel: Iii discussion of results.
\newblock {\em Proceedings of the Physical Society. Section B}, 64(9):747,
  1951.

\bibitem{petch1953cleavage}
NJ~Petch.
\newblock The cleavage strength of polycrystals.
\newblock {\em Journal of the Iron and Steel Institute}, 174:25--28, 1953.

\bibitem{fleck1994strain}
NA~Fleck, GM~Muller, Mike~F Ashby, and John~W Hutchinson.
\newblock Strain gradient plasticity: theory and experiment.
\newblock {\em Acta Metallurgica et materialia}, 42(2):475--487, 1994.

\bibitem{stolken1998microbend}
JS~St{\"o}lken and AG~Evans.
\newblock A microbend test method for measuring the plasticity length scale.
\newblock {\em Acta Materialia}, 46(14):5109--5115, 1998.

\bibitem{uchic2004sample}
Michael~D Uchic, Dennis~M Dimiduk, Jeffrey~N Florando, and William~D Nix.
\newblock Sample dimensions influence strength and crystal plasticity.
\newblock {\em Science}, 305(5686):986--989, 2004.

\bibitem{elmustafa2003nanoindentation}
AA~Elmustafa and DS~Stone.
\newblock Nanoindentation and the indentation size effect: Kinetics of
  deformation and strain gradient plasticity.
\newblock {\em Journal of the Mechanics and Physics of Solids}, 51(2):357--381,
  2003.

\bibitem{zhao2003material}
Minhua Zhao, William~S Slaughter, Ming Li, and Scott~X Mao.
\newblock Material-length-scale-controlled nanoindentation size effects due to
  strain-gradient plasticity.
\newblock {\em Acta Materialia}, 51(15):4461--4469, 2003.

\bibitem{mu2014thickness}
Yang Mu, Ke~Chen, and WJ~Meng.
\newblock Thickness dependence of flow stress of cu thin films in confined
  shear plastic flow.
\newblock {\em MRS Communications}, 4(3):129--133, 2014.

\bibitem{mu2017measuring}
Yang Mu, Xiaoman Zhang, John~W Hutchinson, and Wen~Jin Meng.
\newblock Measuring critical stress for shear failure of interfacial regions in
  coating/interlayer/substrate systems through a micro-pillar testing protocol.
\newblock {\em Journal of Materials Research}, 32(8):1421--1431, 2017.

\bibitem{geers2006size}
Mark~GD Geers, WAM Brekelmans, and PJM Janssen.
\newblock Size effects in miniaturized polycrystalline fcc samples:
  Strengthening versus weakening.
\newblock {\em International Journal of Solids and Structures},
  43(24):7304--7321, 2006.

\bibitem{greer2011plasticity}
Julia~R Greer and Jeff Th~M De~Hosson.
\newblock Plasticity in small-sized metallic systems: Intrinsic versus
  extrinsic size effect.
\newblock {\em Progress in Materials Science}, 56(6):654--724, 2011.

\bibitem{arsenlis1999crystallographic}
A~Arsenlis and DM~Parks.
\newblock Crystallographic aspects of geometrically-necessary and
  statistically-stored dislocation density.
\newblock {\em Acta materialia}, 47(5):1597--1611, 1999.

\bibitem{gao1999mechanism}
H~Gao, Y~Huang, WD~Nix, and JW~Hutchinson.
\newblock Mechanism-based strain gradient plasticity—i. theory.
\newblock {\em Journal of the Mechanics and Physics of Solids},
  47(6):1239--1263, 1999.

\bibitem{acharya2000lattice}
A~Acharya and JL17611250963 Bassani.
\newblock Lattice incompatibility and a gradient theory of crystal plasticity.
\newblock {\em Journal of the Mechanics and Physics of Solids},
  48(8):1565--1595, 2000.

\bibitem{gurtin2002gradient}
Morton~E Gurtin.
\newblock A gradient theory of single-crystal viscoplasticity that accounts for
  geometrically necessary dislocations.
\newblock {\em Journal of the Mechanics and Physics of Solids}, 50(1):5--32,
  2002.

\bibitem{acharya2006size}
Amit Acharya and Anish Roy.
\newblock Size effects and idealized dislocation microstructure at small
  scales: predictions of a phenomenological model of mesoscopic field
  dislocation mechanics: Part i.
\newblock {\em Journal of the Mechanics and Physics of Solids},
  54(8):1687--1710, 2006.

\bibitem{geers2006second}
MGD Geers, WAM Brekelmans, and CJ~Bayley.
\newblock Second-order crystal plasticity: internal stress effects and cyclic
  loading.
\newblock {\em Modelling and Simulation in Materials Science and Engineering},
  15(1):S133, 2006.

\bibitem{han2007finite}
Chung-Souk Han, Anxin Ma, Franz Roters, and Dierk Raabe.
\newblock A finite element approach with patch projection for strain gradient
  plasticity formulations.
\newblock {\em International Journal of Plasticity}, 23(4):690--710, 2007.

\bibitem{kysar2007high}
Jeffrey~W Kysar, Yong~X Gan, Timothy~L Morse, Xi~Chen, and Milton~E Jones.
\newblock High strain gradient plasticity associated with wedge indentation
  into face-centered cubic single crystals: geometrically necessary dislocation
  densities.
\newblock {\em Journal of the Mechanics and Physics of Solids},
  55(7):1554--1573, 2007.

\bibitem{guruprasad2008phenomenological}
PJ~Guruprasad and AA~Benzerga.
\newblock A phenomenological model of size-dependent hardening in crystal
  plasticity.
\newblock {\em Philosophical Magazine}, 88(30-32):3585--3601, 2008.

\bibitem{lele2008small}
SP~Lele and L~Anand.
\newblock A small-deformation strain-gradient theory for isotropic viscoplastic
  materials.
\newblock {\em Philosophical Magazine}, 88(30-32):3655--3689, 2008.

\bibitem{mayeur2011dislocation}
Jason~R Mayeur, David~L McDowell, and Douglas~J Bammann.
\newblock Dislocation-based micropolar single crystal plasticity: Comparison of
  multi-and single criterion theories.
\newblock {\em Journal of the Mechanics and Physics of Solids}, 59(2):398--422,
  2011.

\bibitem{dunne2012crystal}
FPE Dunne, R~Kiwanuka, and AJ~Wilkinson.
\newblock Crystal plasticity analysis of micro-deformation, lattice rotation
  and geometrically necessary dislocation density.
\newblock {\em Proceedings of the Royal Society A: Mathematical, Physical and
  Engineering Sciences}, 468(2145):2509--2531, 2012.

\bibitem{maziere2015strain}
Matthieu Mazi{\`e}re and Samuel Forest.
\newblock Strain gradient plasticity modeling and finite element simulation of
  l{\"u}ders band formation and propagation.
\newblock {\em Continuum Mechanics and Thermodynamics}, 27(1):83--104, 2015.

\bibitem{martinez2016fracture}
Emilio Mart{\'\i}nez-Pa{\~n}eda and Christian~F Niordson.
\newblock On fracture in finite strain gradient plasticity.
\newblock {\em International Journal of Plasticity}, 80:154--167, 2016.

\bibitem{demiral2017enhanced}
Murat Demiral, Kai Nowag, Anish Roy, Rudy Ghisleni, Johann Michler, and Vadim~V
  Silberschmidt.
\newblock Enhanced gradient crystal-plasticity study of size effects in a
  $\beta$-titanium alloy.
\newblock {\em Modelling and Simulation in Materials Science and Engineering},
  25(3):035013, 2017.

\bibitem{gudmundson2019isotropic}
Peter Gudmundson and Carl~FO Dahlberg.
\newblock Isotropic strain gradient plasticity model based on self-energies of
  dislocations and the taylor model for plastic dissipation.
\newblock {\em International Journal of Plasticity}, 121:1--20, 2019.

\bibitem{kuroda2019simple}
Mitsutoshi Kuroda and Alan Needleman.
\newblock A simple model for size effects in constrained shear.
\newblock {\em Extreme Mechanics Letters}, 33:100581, 2019.

\bibitem{zhou2020predictive}
G~Zhou, W~Jeong, ER~Homer, DT~Fullwood, MG~Lee, JH~Kim, H~Lim, H~Zbib, and
  RH~Wagoner.
\newblock A predictive strain-gradient model with no undetermined constants or
  length scales.
\newblock {\em Journal of the Mechanics and Physics of Solids}, 145:104178,
  2020.

\bibitem{sun2019strain}
Fengwei Sun, Edward~D Meade, and Noel~P O'Dowd.
\newblock Strain gradient crystal plasticity modelling of size effects in a
  hierarchical martensitic steel using the voronoi tessellation method.
\newblock {\em International Journal of Plasticity}, 119:215--229, 2019.

\bibitem{haouala2020simulation}
S~Haouala, S~Lucarini, J~LLorca, and J~Segurado.
\newblock Simulation of the hall-petch effect in fcc polycrystals by means of
  strain gradient crystal plasticity and fft homogenization.
\newblock {\em Journal of the Mechanics and Physics of Solids}, 134:103755,
  2020.

\bibitem{berbenni2020fast}
St{\'e}phane Berbenni, Vincent Taupin, and Ricardo~A Lebensohn.
\newblock A fast fourier transform-based mesoscale field dislocation mechanics
  study of grain size effects and reversible plasticity in polycrystals.
\newblock {\em Journal of the Mechanics and Physics of Solids}, 135:103808,
  2020.

\bibitem{mayeur2014comparison}
JR~Mayeur and DL~McDowell.
\newblock A comparison of gurtin type and micropolar theories of generalized
  single crystal plasticity.
\newblock {\em International Journal of Plasticity}, 57:29--51, 2014.

\bibitem{voyiadjis2019strain}
George~Z Voyiadjis and Yooseob Song.
\newblock Strain gradient continuum plasticity theories: theoretical, numerical
  and experimental investigations.
\newblock {\em International Journal of Plasticity}, 121:21--75, 2019.

\bibitem{evers2004non}
LP~Evers, WAM Brekelmans, and MGD Geers.
\newblock Non-local crystal plasticity model with intrinsic ssd and gnd
  effects.
\newblock {\em Journal of the Mechanics and Physics of Solids},
  52(10):2379--2401, 2004.

\bibitem{yefimov2005size}
S~Yefimov and E~Van~der Giessen.
\newblock Size effects in single crystal thin films: nonlocal crystal
  plasticity simulations.
\newblock {\em European Journal of Mechanics-A/Solids}, 24(2):183--193, 2005.

\bibitem{bayley2006comparison}
CJ~Bayley, WAM Brekelmans, and MGD Geers.
\newblock A comparison of dislocation induced back stress formulations in
  strain gradient crystal plasticity.
\newblock {\em International Journal of Solids and Structures},
  43(24):7268--7286, 2006.

\bibitem{kuroda2006studies}
Mitsutoshi Kuroda and Viggo Tvergaard.
\newblock Studies of scale dependent crystal viscoplasticity models.
\newblock {\em Journal of the Mechanics and Physics of Solids},
  54(9):1789--1810, 2006.

\bibitem{forest2008some}
Samuel Forest.
\newblock Some links between cosserat, strain gradient crystal plasticity and
  the statistical theory of dislocations.
\newblock {\em Philosophical Magazine}, 88(30-32):3549--3563, 2008.

\bibitem{kapoor2018incorporating}
Kartik Kapoor, Yung Suk~Jeremy Yoo, Todd~A Book, Josh~P Kacher, and Michael~D
  Sangid.
\newblock Incorporating grain-level residual stresses and validating a crystal
  plasticity model of a two-phase ti-6al-4 v alloy produced via additive
  manufacturing.
\newblock {\em Journal of the Mechanics and Physics of Solids}, 121:447--462,
  2018.

\bibitem{bandyopadhyay2021comparative}
Ritwik Bandyopadhyay, Sven~E Gustafson, Kartik Kapoor, Diwakar Naragani,
  Darren~C Pagan, and Michael~D Sangid.
\newblock Comparative assessment of backstress models using high-energy x-ray
  diffraction microscopy experiments and crystal plasticity finite element
  simulations.
\newblock {\em International Journal of Plasticity}, 136:102887, 2021.

\bibitem{taylor1934mechanism}
Geoffrey~Ingram Taylor.
\newblock The mechanism of plastic deformation of crystals. part
  i.—theoretical.
\newblock {\em Proceedings of the Royal Society of London. Series A, Containing
  Papers of a Mathematical and Physical Character}, 145(855):362--387, 1934.

\bibitem{khan1995continuum}
Akhtar~S Khan and Sujian Huang.
\newblock {\em Continuum theory of plasticity}.
\newblock John Wiley \& Sons, 1995.

\bibitem{weber1990finite}
Gustavo Weber and Lallit Anand.
\newblock Finite deformation constitutive equations and a time integration
  procedure for isotropic, hyperelastic-viscoplastic solids.
\newblock {\em Computer Methods in Applied Mechanics and Engineering},
  79(2):173--202, 1990.

\bibitem{kocks1975thermodynamics}
Ulrich~Fred Kocks, A~Argon, and M~Ashby.
\newblock Thermodynamics and kinetics of slip.
\newblock 1975.

\bibitem{mishra2019new}
Sumeet Mishra, Manasij Yadava, Kaustubh~N Kulkarni, and NP~Gurao.
\newblock A new phenomenological approach for modeling strain hardening
  behavior of face centered cubic materials.
\newblock {\em Acta Materialia}, 178:99--113, 2019.

\bibitem{estrin1996dislocation}
Yuri Estrin.
\newblock Dislocation-density-related constitutive modeling.
\newblock {\em Unified constitutive laws of plastic deformation}, 1:69--106,
  1996.

\bibitem{kocks2003physics}
UF~Kocks and H~Mecking.
\newblock Physics and phenomenology of strain hardening: the fcc case.
\newblock {\em Progress in materials science}, 48(3):171--273, 2003.

\bibitem{dai1997geometrically}
Hong Dai.
\newblock {\em Geometrically-necessary dislocation density in continuum
  plasticity theory, FEM implementation and applications}.
\newblock PhD thesis, Massachusetts Institute of Technology, 1997.

\bibitem{nix1998indentation}
William~D Nix and Huajian Gao.
\newblock Indentation size effects in crystalline materials: a law for strain
  gradient plasticity.
\newblock {\em Journal of the Mechanics and Physics of Solids}, 46(3):411--425,
  1998.

\bibitem{evers2004scale}
LP~Evers, WAM Brekelmans, and MGD Geers.
\newblock Scale dependent crystal plasticity framework with dislocation density
  and grain boundary effects.
\newblock {\em International Journal of solids and structures},
  41(18-19):5209--5230, 2004.

\bibitem{cuitino1993computational}
Alberto~M Cuitino and Michael Ortiz.
\newblock Computational modelling of single crystals.
\newblock {\em Modelling and Simulation in Materials Science and Engineering},
  1(3):225, 1993.

\bibitem{mcginty2001multiscale}
Robert~Davis McGinty.
\newblock {\em Multiscale representation of polycrystalline inelasticity}.
\newblock PhD thesis, Georgia institute of technology, 2001.

\bibitem{mcginty2006semi}
RD~McGinty and DL~McDowell.
\newblock A semi-implicit integration scheme for rate independent finite
  crystal plasticity.
\newblock {\em International Journal of Plasticity}, 22(6):996--1025, 2006.

\bibitem{ling2005numerical}
Xianwu Ling, MF~Horstemeyer, and GP~Potirniche.
\newblock On the numerical implementation of 3d rate-dependent single crystal
  plasticity formulations.
\newblock {\em International Journal for Numerical Methods in Engineering},
  63(4):548--568, 2005.

\bibitem{curtis1990algebraic}
Morton~L Curtis.
\newblock Algebraic preliminaries.
\newblock In {\em Abstract Linear Algebra}, pages 1--7. Springer, 1990.

\bibitem{petersen2008matrix}
Kaare~Brandt Petersen, Michael~Syskind Pedersen, et~al.
\newblock The matrix cookbook.
\newblock {\em Technical University of Denmark}, 7(15):510, 2008.

\bibitem{permann2020moose}
Cody~J Permann, Derek~R Gaston, David Andr{\v{s}}, Robert~W Carlsen, Fande
  Kong, Alexander~D Lindsay, Jason~M Miller, John~W Peterson, Andrew~E
  Slaughter, Roy~H Stogner, et~al.
\newblock Moose: Enabling massively parallel multiphysics simulation.
\newblock {\em SoftwareX}, 11:100430, 2020.

\bibitem{kothari1998elasto}
M~Kothari and L~Anand.
\newblock Elasto-viscoplastic constitutive equations for polycrystalline
  metals: application to tantalum.
\newblock {\em Journal of the Mechanics and Physics of Solids}, 46(1):51--83,
  1998.

\bibitem{ranjan2021crystal}
Devraj Ranjan, Sankar Narayanan, Kai Kadau, and Anirban Patra.
\newblock Crystal plasticity modeling of non-schmid yield behavior: from ni3al
  single crystals to ni-based superalloys.
\newblock {\em Modelling and Simulation in Materials Science and Engineering},
  29(5):055005, 2021.

\bibitem{sedighiani2021determination}
Karo Sedighiani, Konstantina Traka, Franz Roters, Dierk Raabe, Jilt Sietsma,
  and Martin Diehl.
\newblock Determination and analysis of the constitutive parameters of
  temperature-dependent dislocation-density-based crystal plasticity models.
\newblock {\em Mechanics of Materials}, page 104117, 2021.

\bibitem{quey2011large}
Romain Quey, PR~Dawson, and Fabrice Barbe.
\newblock Large-scale 3d random polycrystals for the finite element method:
  Generation, meshing and remeshing.
\newblock {\em Computer Methods in Applied Mechanics and Engineering},
  200(17-20):1729--1745, 2011.

\bibitem{trelisadvanced}
Csimsoft Trelis.
\newblock Advanced meshing for challenging simulations (2013-2016).

\bibitem{cordero2016six}
Zachary~C Cordero, Braden~E Knight, and Christopher~A Schuh.
\newblock Six decades of the hall--petch effect--a survey of grain-size
  strengthening studies on pure metals.
\newblock {\em International Materials Reviews}, 61(8):495--512, 2016.

\bibitem{mishra2009widths}
SK~Mishra, P~Pant, K~Narasimhan, AD~Rollett, and I~Samajdar.
\newblock On the widths of orientation gradient zones adjacent to grain
  boundaries.
\newblock {\em Scripta materialia}, 61(3):273--276, 2009.

\bibitem{el2015unravelling}
Jaafar~A El-Awady.
\newblock Unravelling the physics of size-dependent dislocation-mediated
  plasticity.
\newblock {\em Nature communications}, 6(1):1--9, 2015.

\bibitem{argon2008strengthening}
Ali Argon.
\newblock {\em Strengthening mechanisms in crystal plasticity}, volume~4.
\newblock Oxford University Press on Demand, 2008.

\bibitem{ono1982grain}
Noboru Ono and Seiichi Karashima.
\newblock Grain size dependence of flow stress in copper polycrystals.
\newblock {\em Scripta Metallurgica}, 16(4):381--384, 1982.

\bibitem{hansen1977effect}
Niels Hansen.
\newblock The effect of grain size and strain on the tensile flow stress of
  aluminium at room temperature.
\newblock {\em Acta Metallurgica}, 25(8):863--869, 1977.

\bibitem{eisenlohr2013spectral}
Philip Eisenlohr, Martin Diehl, Ricardo~A Lebensohn, and Franz Roters.
\newblock A spectral method solution to crystal elasto-viscoplasticity at
  finite strains.
\newblock {\em International Journal of Plasticity}, 46:37--53, 2013.

\end{thebibliography}

\end{document}